\documentclass{article} 
\usepackage{iclr2022_conference,times}

\usepackage{hyperref}
\usepackage{url}

\usepackage{amssymb,amsmath,amsthm,enumerate,threeparttable,bm,graphicx,subfigure}
\usepackage{algorithm, algorithmic}
\usepackage{booktabs}
\usepackage{multirow}

\usepackage{color}
\def\red#1{\textcolor{red}{#1}}

\newcommand{\tabincell}[2]{\begin{tabular}{@{}c#1@{}}#2\end{tabular}} 

\usepackage{enumitem}
\setenumerate[1]{itemsep=0pt,partopsep=0pt,parsep=\parskip,topsep=5pt}
\setitemize[1]{itemsep=0pt,partopsep=0pt,parsep=\parskip,topsep=5pt}
\setdescription{itemsep=0pt, partopsep=0pt,parsep=\parskip,topsep=5pt}

\long\def\comment#1{}

\title{Backdoor Defense via Decoupling the Training Process}

\author{Kunzhe Huang$^{1,}$\thanks{The first two authors contributed equally to this work. This work was mostly done when Kunzhe Huang and Yiming Li were the research interns at The Chinese University of Hong Kong, Shenzhen. $\dagger$ indicates corresponding authors: Baoyuan Wu (\href{mailto:wubaoyuan@cuhk.edu.cn}{wubaoyuan@cuhk.edu.cn}) and Zhan Qin (\href{mailto:qinzhan@zju.edu.cn}{qinzhan@zju.edu.cn}).} , Yiming Li$^{3,\ast}$, Baoyuan Wu$^{2, \dagger}$, Zhan Qin$^{1, \dagger}$, Kui Ren$^{1}$\\
$^{1}$School of Cyber Science and Technology, Zhejiang University\\
$^{2}$School of Data Science, Shenzhen Research Institute of Big Data, The Chinese University of Hong Kong, Shenzhen\\
$^{3}$Tsinghua Shenzhen International Graduate School, Tsinghua University\\
\{hkunzhe, zhanqin, kuiren\}@zju.edu.cn; wubaoyuan@cuhk.edu.cn; li-ym18@mails.tsinghua.edu.cn
}

\iclrfinalcopy 
\begin{document}

\maketitle

\begin{abstract}
Recent studies have revealed that deep neural networks (DNNs) are vulnerable to backdoor attacks, where attackers embed hidden backdoors in the DNN model by poisoning a few training samples. The attacked model behaves normally on benign samples, whereas its prediction will be maliciously changed when the backdoor is activated. We reveal that poisoned samples tend to cluster together in the feature space of the attacked DNN model, which is mostly due to the end-to-end supervised training paradigm. Inspired by this observation, we propose a novel backdoor defense via decoupling the original end-to-end training process into three stages. Specifically, we first learn the backbone of a DNN model via \emph{self-supervised learning} based on training samples without their labels. The learned backbone will map samples with the same ground-truth label to similar locations in the feature space. Then, we freeze the parameters of the learned backbone and train the remaining fully connected layers via standard training with all (labeled) training samples. Lastly, to further alleviate side-effects of poisoned samples in the second stage, we remove labels of some `low-credible' samples determined based on the learned model and conduct a \emph{semi-supervised fine-tuning} of the whole model. Extensive experiments on multiple benchmark datasets and DNN models verify that the proposed defense is effective in reducing backdoor threats while preserving high accuracy in predicting benign samples. Our code is available at \url{https://github.com/SCLBD/DBD}.
\end{abstract}

\section{Introduction}
\label{sec:intro}

\comment{
\begin{figure}[ht]
    \centering
    \includegraphics[width=0.473\textwidth]{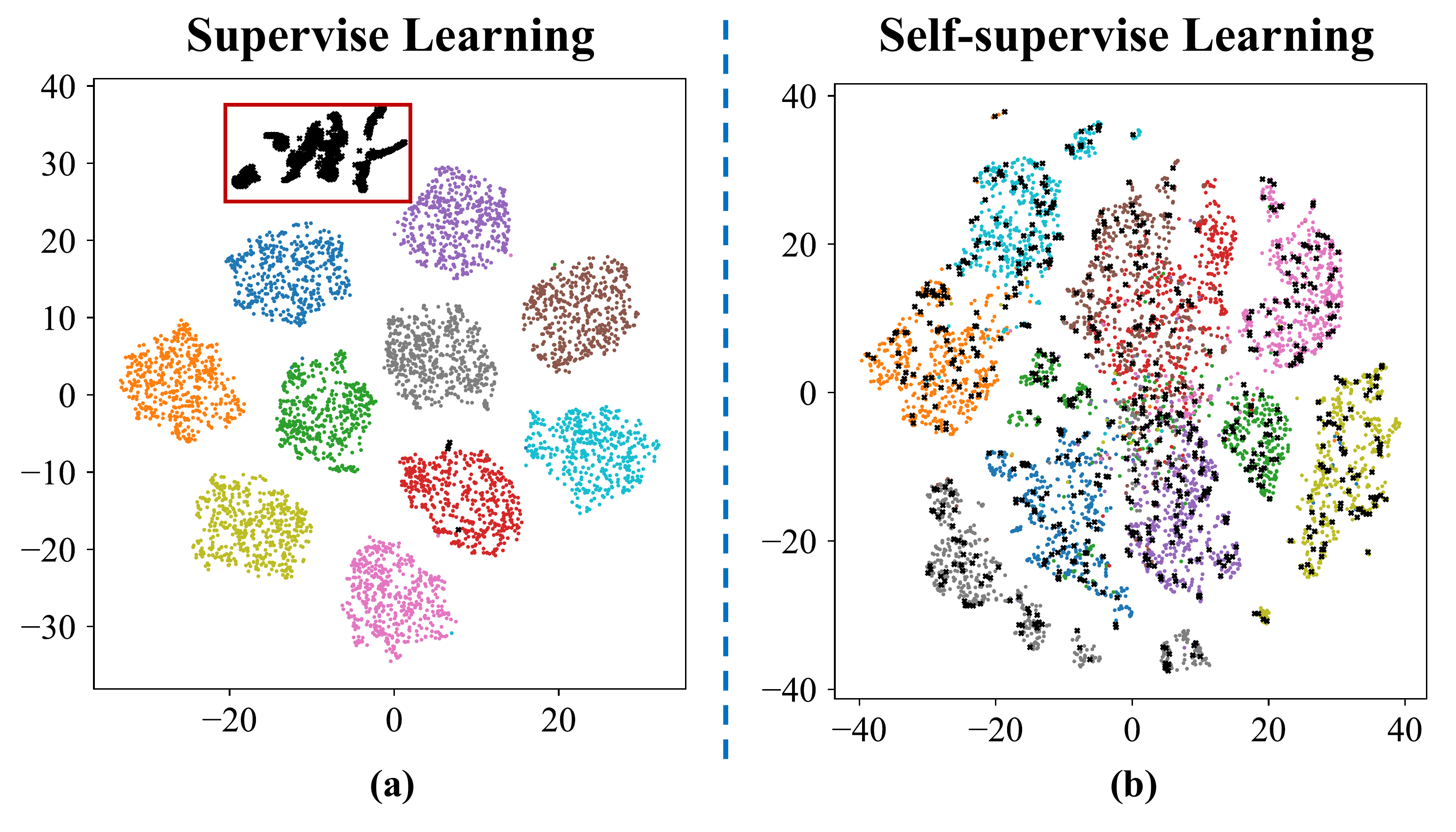}
    \caption{The t-SNE of poisoned samples in the hidden space generated by attacked DNNs trained in supervised and self-supervised manner. As shown in the figure, poisoned samples tend to cluster together of DNN (within the red box) trained with standard supervised learning, whereas lie closely to samples with their ground-truth label of the one trained with self-supervised learning. }
    \label{fig:intro}
    \vspace{-1em}
\end{figure}
}

Deep learning, especially deep neural networks (DNNs), has been widely adopted in many realms \citep{wang2020pillar,li2020short,wen20adaptive} for its high effectiveness. In general, the training of DNNs requires a large amount of training samples and computational resources. Accordingly, third-party resources ($e.g.$, third-party data or servers) are usually involved. While the opacity of the training process brings certain convenience, it also introduces new security threats.

Backdoor attack poses a new security threat to the training process of DNNs \citep{li2020backdoor}. It maliciously manipulates the prediction of the attacked DNNs by poisoning a few training samples. Specifically, backdoor attackers inject the \emph{backdoor trigger} ($i.e.$, a particular pattern) to some benign training images and change their labels with the attacker-specified \emph{target label}. The connection between the backdoor trigger and the target label will be learned by DNNs during the training process. In the inference process, the prediction of attacked DNNs will be changed to the target label when the trigger is present, whereas the attacked DNNs will behave normally on benign samples. As such, users are difficult to realize the existence of hidden backdoors and therefore this attack is a serious threat to the practical applications of DNNs.

In this paper, we first investigate backdoor attacks from the hidden feature space. Our preliminary experiments reveal that the backdoor is embedded in the feature space, $i.e.$, samples with the backdoor trigger (dubbed \emph{poisoned samples}) tend to cluster together in the feature space. We reveal that \emph{this phenomenon is mostly due to the end-to-end supervised training paradigm}. Specifically, the excessive learning capability allows DNNs to learn features about the backdoor trigger, while the DNNs can shrink the distance between poisoned samples in the feature space and connect the learned trigger-related features with the target label by the end-to-end supervised training. Based on this understanding, we propose to decouple the end-to-end training process for the backdoor defense. Specifically, we treat the DNNs as two disjoint parts, including a \emph{feature extractor} ($i.e.$, backbone) and a \emph{simple classifier} ($i.e.$, the remaining fully connected layers). We first learn the \emph{purified feature extractor} via \emph{self-supervised learning} \citep{kolesnikov2019revisiting,chen2020simple,jing2020self} with unlabeled training samples (obtained by removing their labels), and then learn the simple classifier via standard supervised training process based on the learned feature extractor and all training samples. The strong data augmentations involved in the self-supervised learning damage trigger patterns, making them unlearnable during representation learning; and the decoupling process further disconnects trigger patterns and the target label. Accordingly, hidden backdoors cannot be successfully created even the model is trained on the poisoned dataset based on our defense.

Moreover, we further reveal that the representation of poisoned samples generated by the purified extractor is significantly different from those generated by the extractor learned with standard training process. Specifically, the poisoned sample lies closely to samples with its ground-truth label instead of the target label. This phenomenon makes the training of the simple classifier similar to \emph{label-noise learning} \citep{wang2019symmetric,ma2020normalized,berthon2021confidence}. As such, we first filter \emph{high-credible training samples} ($i.e.$, training samples that are most probably to be benign) and then use those samples as labeled samples and the remaining part to form unlabeled samples to fine-tune the whole model via \emph{semi-supervised learning} \citep{rasmus2015semi,berthelot2019mixmatch,sohn2020fixmatch}. This approach is to further reduce the adverse effects of poisoned samples.

The main contributions of this paper are three-fold. \textbf{(1)} We reveal that the backdoor is embedded in the feature space, which is mostly due to the end-to-end supervised training paradigm. \textbf{(2)} Based on our understanding, we propose a decoupling-based backdoor defense (DBD) to alleviate the threat of poisoning-based backdoor attacks. \textbf{(3)} Experiments on classical benchmark datasets are conducted, which verify the effectiveness of our defense.

\vspace{-0.5em}
\section{Related Work}
\vspace{-0.5em}

\subsection{Backdoor Attack}
\label{sec:related_attack}

Backdoor attack is an emerging research area, which raises security concerns about training with third-party resources. In this paper, we focus on the poisoning-based backdoor attack towards image classification, where attackers can only modify the dataset instead of other training components ($e.g.$, training loss). This threat could also happen in other tasks \citep{xiang2021backdoor,zhai2021backdoor,li2022few} and with different attacker's capacities \citep{nguyen2020input,tang2020embarrassingly,zeng2021rethinking}, which are out-of-scope of this paper. 
In general, existing attacks can be divided into two main categories based on the property of target labels, as follows:

\noindent \textbf{Poison-Label Backdoor Attack.}
It is currently the most common attack paradigm, where the target label is different from the ground-truth label of poisoned samples. BadNets \citep{gu2017badnets} is the first and most representative poison-label attack. 
Specifically, it randomly selected a few samples from the original benign dataset to generate \emph{poisoned samples} by stamping the \emph{backdoor trigger} onto the (benign) image and change their label with an attacker-specified \emph{target label}. Those generated poisoned samples associated with remaining benign ones were combined to form the \emph{poisoned training dataset}, which will be delivered to users. After that, \citep{chen2017targeted} suggested that the poisoned image should be similar to its benign version for the stealthiness, based on which they proposed the \emph{blended attack}. 
Recently, \citep{xue2020one,li2020invisible,li2021invisible} further explored how to conduct poison-label backdoor attacks more stealthily. Most recently, a more stealthy and effective attack, the WaNet \citep{nguyen2021wanet}, was proposed. WaNet adopted image warping as the backdoor trigger, which deforms but preserves the image content.

\noindent \textbf{Clean-Label Backdoor Attack.} 
Although the poisoned image generated by poison-label attacks could be similar to its benign version, users may still notice the attack by examining the image-label relationship. To address this problem, \cite{turner2019label} proposed the \emph{clean-label attack paradigm}, where the target label is consistent with the ground-truth label of poisoned samples. Specifically, they first leveraged adversarial perturbations or generative models to modify some benign images from the target class and then conducted the standard trigger injection process. 
This idea was generalized to attack video classification in \citep{zhao2020clean}, where they adopted the targeted universal adversarial perturbation \citep{moosavi2017universal} as the trigger pattern. 
Although clean-label backdoor attacks are more stealthy compared with poison-label ones, they usually suffer from relatively poor performance and may even fail in creating backdoors \citep{li2020backdoor}.

\subsection{Backdoor Defense}
Currently, there are also some approaches to alleviate the backdoor threat. Existing defenses are mostly \emph{empirical}, which can be divided into five main categories, including \textbf{(1)} \emph{detection-based defenses} \citep{xu2021detecting,zeng2021rethinking,xiang2022post}, \textbf{(2)} \emph{preprocessing based defenses} \citep{doan2020februus,li2021backdoor2,zeng2021deepsweep}, \textbf{(3)} \emph{model reconstruction based defenses} \citep{zhao2020bridging,li2021neural,zeng2022adversarial}, \textbf{(4)} \emph{trigger synthesis based defenses} \citep{guo2020towards,dong2021black,shen2021backdoor}, and \textbf{(5)} \emph{poison suppression based defenses} \citep{du2019robust,borgnia2021strong}. Specifically, detection-based defenses examine whether a suspicious DNN or sample is attacked and it will deny the use of malicious objects; Preprocessing based methods intend to damage trigger patterns contained in attack samples to prevent backdoor activation by introducing a preprocessing module before feeding images into DNNs; Model reconstruction based ones aim at removing the hidden backdoor in DNNs by modifying models directly; The fourth type of defenses synthesize potential trigger patterns at first, following by the second stage that the hidden backdoor is eliminated by suppressing their effects; The last type of methods depress the effectiveness of poisoned samples during the training process to prevent the creation of hidden backdoors. In general, our method is most relevant to this type of defenses. 

In this paper, we only focus on the last four types of defenses since they directly improve the robustness of DNNs. Besides, there were also few works focusing on \emph{certified backdoor defenses} \citep{wang2020certifying,weber2020rab}.
Their robustness is theoretically guaranteed under certain assumptions, which cause these methods to be generally weaker than empirical ones in practice.

\subsection{Semi-supervised and Self-supervised Learning}

\noindent \textbf{Semi-supervised Learning.} 
In many real-world applications, the acquisition of labeled data often relies on manual labeling, which is very expensive. In contrast, obtaining unlabeled samples is much easier. To utilize the power of unlabeled samples with labeled ones simultaneously, a great amount of semi-supervised learning methods were proposed \citep{gao2017semi,berthelot2019mixmatch,van2020survey}. Recently, semi-supervised learning was also introduced in improving the security of DNNs \citep{stanforth2019,carmon2019}, where they utilized unlabelled samples in the adversarial training. Most recently, \citep{yan2021dehib} discussed how to backdoor semi-supervised learning. However, this approach needs to control other training components ($e.g.$, training loss) in addition to modifying training samples and therefore is out-of-scope of this paper. How to adopt semi-supervised learning for backdoor defense remains blank.

\noindent \textbf{Self-supervised Learning.}
This learning paradigm is a subset of \emph{unsupervised learning}, where DNNs are trained with supervised signals generated from the data itself \citep{chen2020simple,grill2020bootstrap,liu2021self}. It has been adopted for increasing adversarial robustness \citep{hendrycks2019using,wu2021adversarial,shi2021online}. Most recently, there were also a few works \citep{saha2021backdoor,carlini2021poisoning,jia2021badencoder} exploring how to backdoor self-supervised learning. However, these attacks are out-of-scope of this paper since they need to control other training components ($e.g.$, training loss) in addition to modifying training samples.

\vspace{-0.5em}
\section{Revisiting Backdoor Attacks from the Hidden Feature Space}
\vspace{-0.5em}
\label{sec:motivation}

\begin{figure}[ht]
\centering
\vspace{-3em}
\subfigure[]{
\centering
\includegraphics[width=0.235\textwidth]{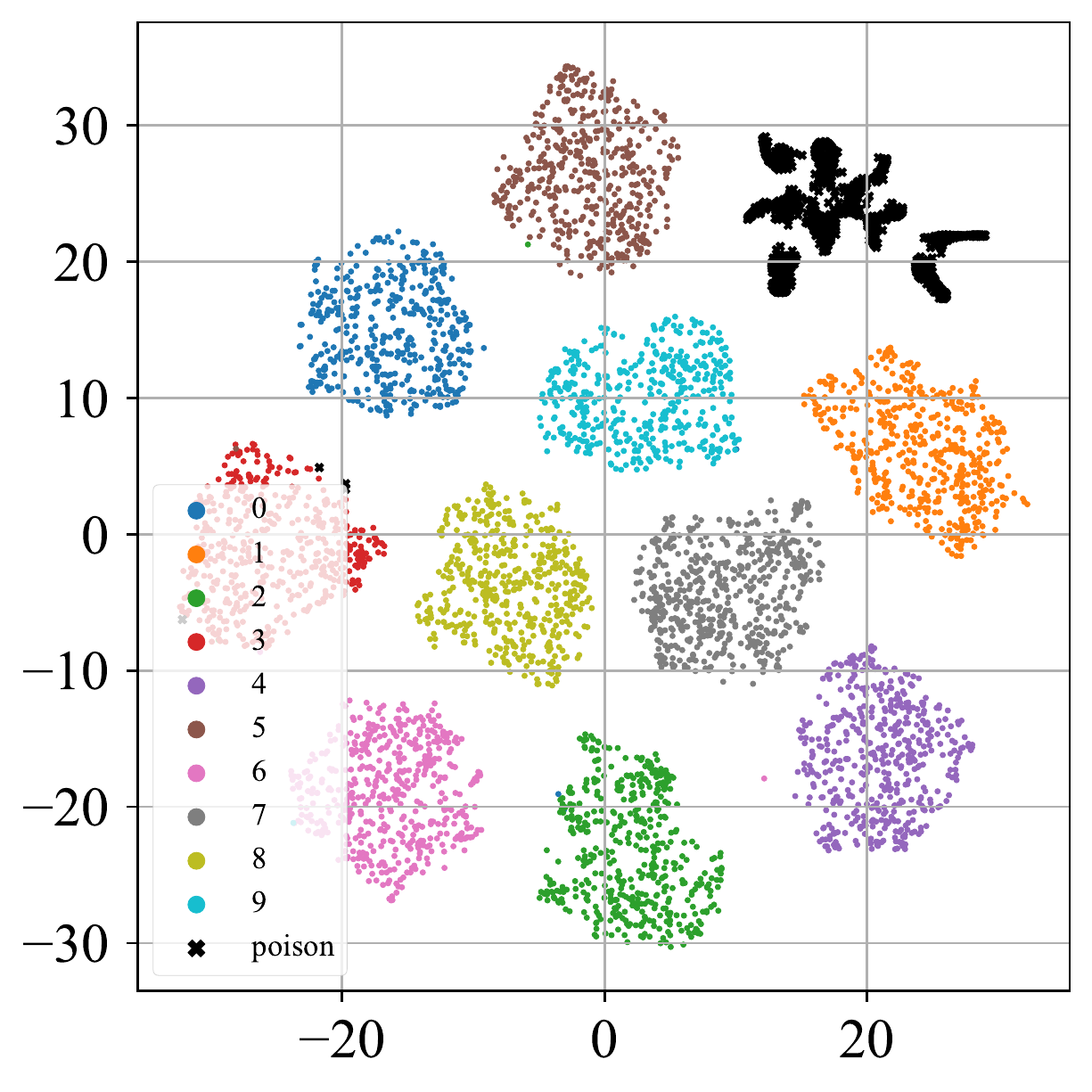}\label{fig:m_a}}
\vspace{-0.8em}
\subfigure[]{
\centering
\includegraphics[width=0.235\textwidth]{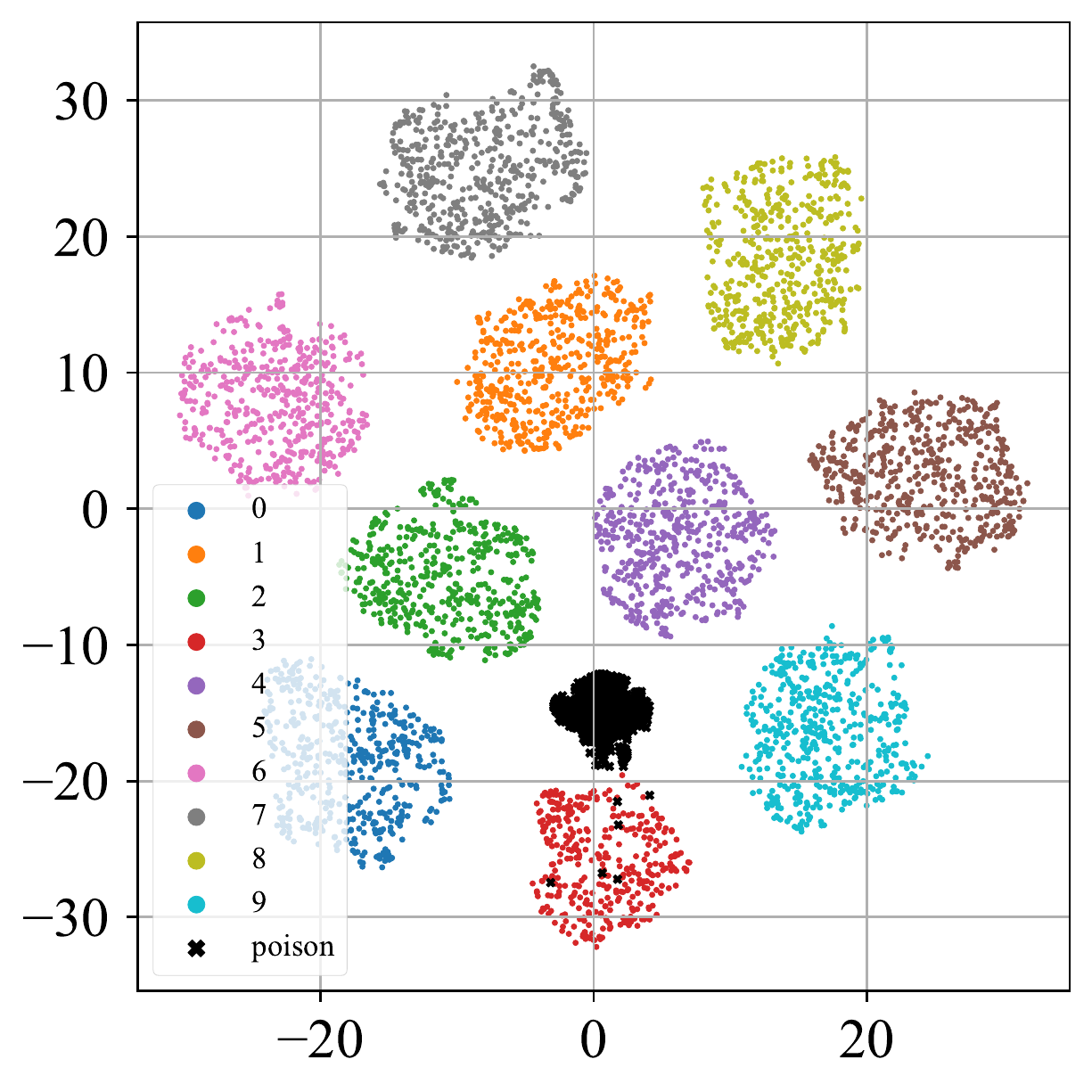}\label{fig:m_b}}
\vspace{-0.8em}
\subfigure[]{
\centering
\includegraphics[width=0.235\textwidth]{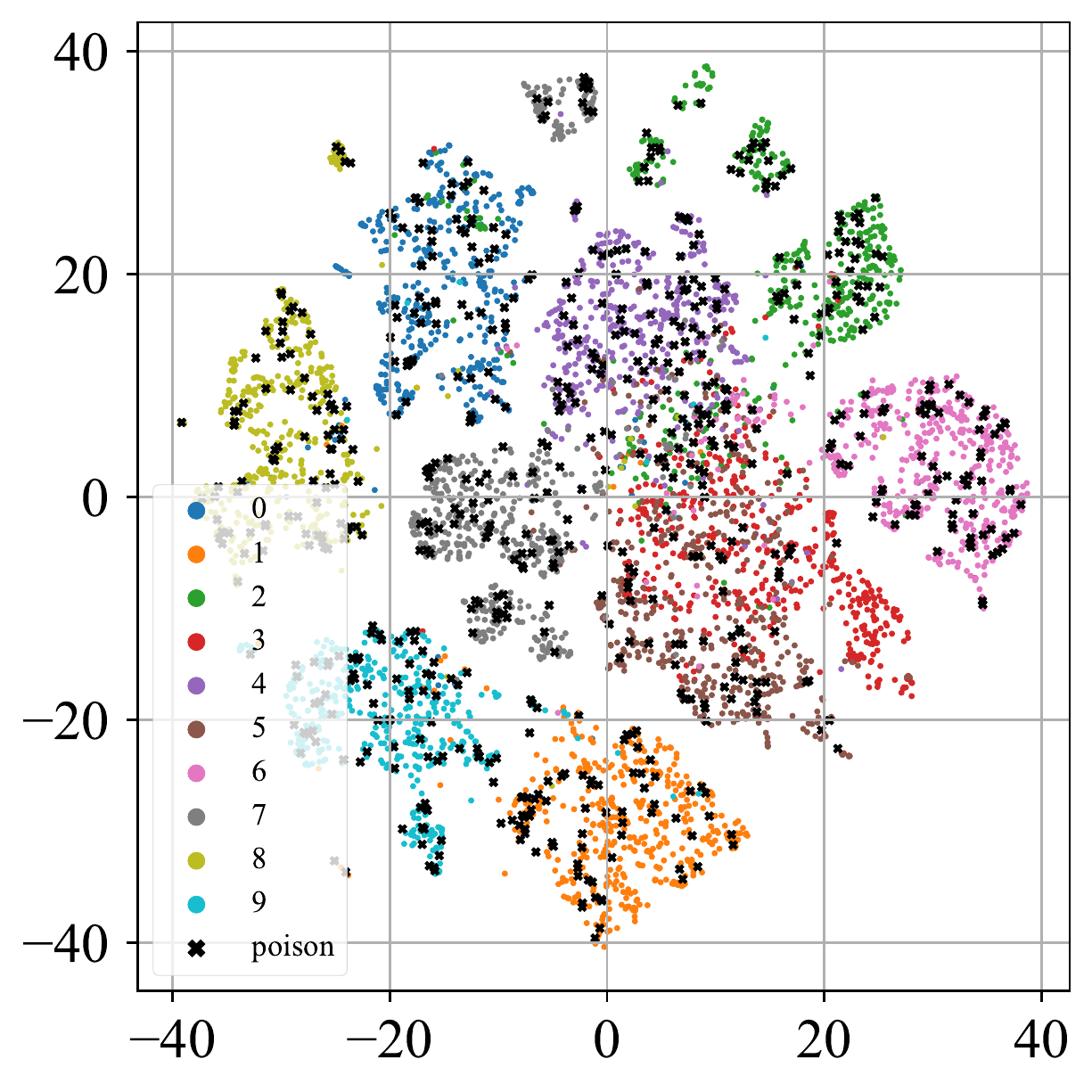}\label{fig:m_c}}
\vspace{0.3em}
\subfigure[]{
\centering
\includegraphics[width=0.235\textwidth]{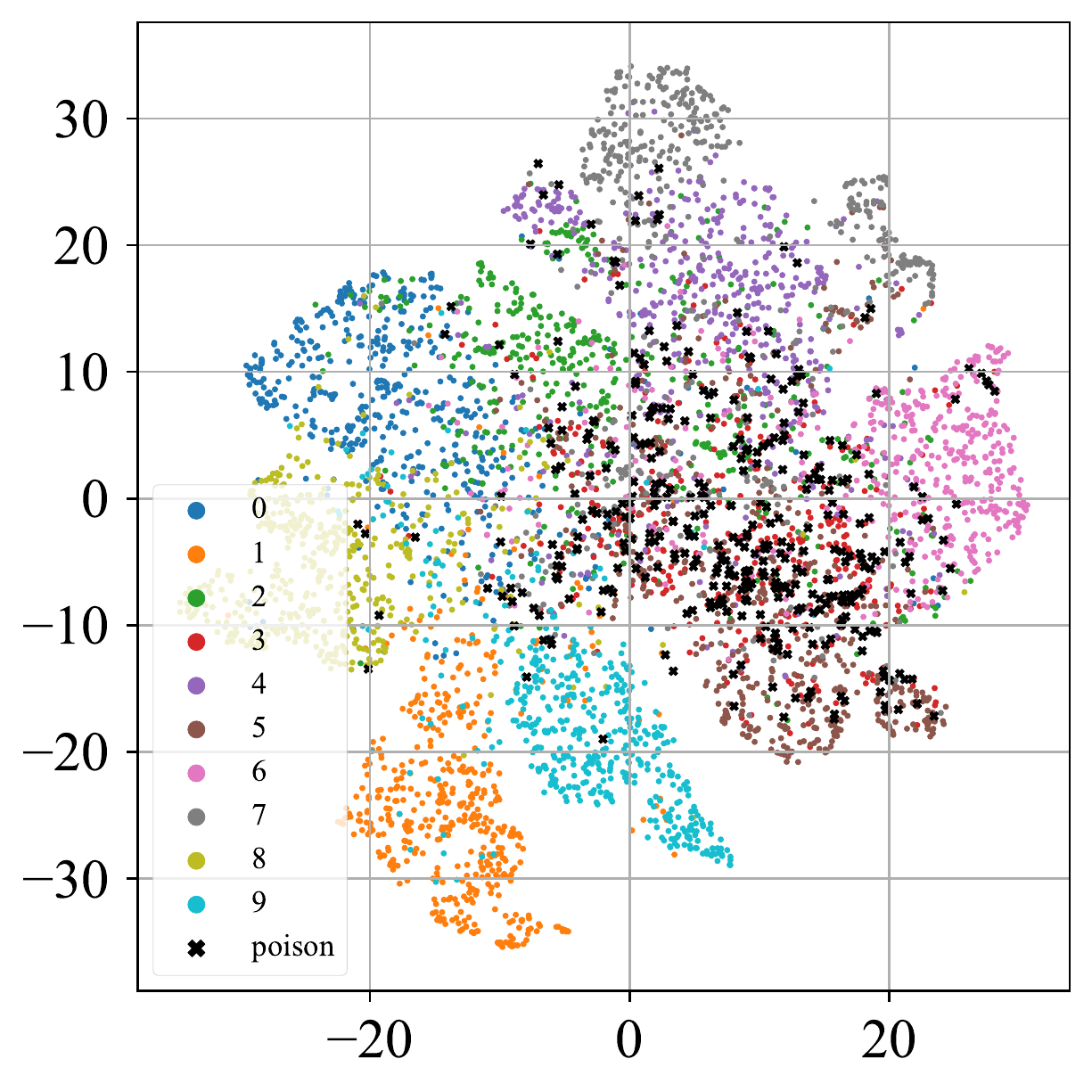}\label{fig:m_d}}
\vspace{0.3em}
\caption{The t-SNE of poisoned samples in the hidden space generated by different models. \textbf{(a)-(b)}: DNNs trained with supervised learning. \textbf{(c)-(d)}: DNNs trained with self-supervised learning. \textbf{(a)\&(c)}: DNNs under BadNets attack. \textbf{(b)\&(d)}: DNNs under label-consistent attack. No matter under the BadNets or label-consistent attack, poisoned samples tend to cluster together in the hidden space generated by DNNs trained with supervised learning, whereas lie closely to samples with their ground-truth label by those trained with self-supervised learning.}
\label{fig:motivation}
\vspace{-1.5em}
\end{figure}

In this section, we analyze the behavior of poisoned samples from the hidden feature space of attacked models and discuss its inherent mechanism.

\noindent \textbf{Settings.} 
We conduct the BadNets \citep{gu2017badnets} and label-consistent attack \citep{turner2019label} on CIFAR-10 dataset \citep{krizhevsky2009learning} for the discussion. They are representative of poison-label attacks and clean-label attacks, respectively. Specifically, we conduct supervised learning on the poisoned datasets with the standard training process and self-supervised learning on the unlabelled poisoned datasets with SimCLR \citep{chen2020simple}. We visualize poisoned samples in the hidden feature space generated by attacked DNNs based on the t-SNE \citep{van2008visualizing}. More detailed settings are presented in Appendix \ref{sec:motivation_set}.

\noindent \textbf{Results.} 
As shown in Figure \ref{fig:m_a}-\ref{fig:m_b}, poisoned samples (denoted by `black-cross') tend to cluster together to form a separate cluster after the standard supervised training process, no matter under the poison-label attack or clean-label attack. This phenomenon implies why existing poisoning-based backdoor attacks can succeed. Specifically, the excessive learning capability allows DNNs to learn features about the backdoor trigger. Associated with the end-to-end supervised training paradigm, DNNs can shrink the distance between poisoned samples in the feature space and connect the learned trigger-related features with the target label. In contrast, as shown in Figure \ref{fig:m_c}-\ref{fig:m_d}, poisoned samples lie closely to samples with their ground-truth label after the self-supervised training process on the unlabelled poisoned dataset. 
It indicates that we can prevent the creation of backdoors by self-supervised learning, which will be further introduced in the next section.

\section{Decoupling-based Backdoor Defense}

\subsection{Preliminaries}
\noindent \textbf{General Pipeline of Backdoor Attacks.} 
Let $\mathcal{D} = \{ (\bm{x}_i, y_i) \}_{i=1}^{N}$ denotes the benign training set, where $\bm{x}_i \in \mathcal{X}= \{0,1,\ldots, 255\}^{C\times W \times H}$ is the image, $y_i \in \mathcal{Y} = \{0,1,\ldots, K\}$ is its label, $K$ is the number of classes, and $y_t \in \mathcal{Y}$ indicates the \emph{target label}. How to generate the poisoned dataset $\mathcal{D}_{p}$ is the cornerstone of backdoor attacks. Specifically, $\mathcal{D}_{p}$ consists of two subsets, including the modified version of a subset of $\mathcal{D}$ and remaining benign samples, $i.e.$, 
$\mathcal{D}_{p} =  \mathcal{D}_{m} \cup \mathcal{D}_{b}$, where $\mathcal{D}_{b} \subset \mathcal{D}$, $\gamma \triangleq \frac{|\mathcal{D}_{m}|}{|\mathcal{D}|}$ is the \emph{poisoning rate}, 
$\mathcal{D}_{m} = \left\{(\bm{x}', y_t)| \bm{x}' = G(\bm{x}), (\bm{x},y) \in \mathcal{D} \backslash \mathcal{D}_{b} \right\}$, and $G: \mathcal{X} \rightarrow \mathcal{X}$ is an attacker-predefined poisoned image generator. For example, $G(\bm{x}) = (\bm{1}-\bm{\lambda}) \otimes \bm{x}+ \bm{\lambda} \otimes \bm{t}$, where $\bm{\lambda} \in [0,1]^{C \times W \times H}$, $\bm{t} \in \mathcal{X}$ is the \emph{trigger} pattern, and $\otimes$ is the element-wise product in the blended attack \citep{chen2017targeted}. Once $\mathcal{D}_{p}$ is generated, it will be sent to users who will train DNNs on it. Hidden backdoors will be created after the training process.

\noindent \textbf{Threat Model.} 
In this paper, we focus on defending against poisoning-based backdoor attacks. The attacker can arbitrarily modify the training set whereas cannot change other training components ($e.g.$, model structure and training loss). For our proposed defense, we assume that defenders can fully control the training process. This is the scenario that users adopt third-party collected samples for training. Note that we do not assume that defenders have a local benign dataset, which is often required in many existing defenses \citep{wang2019neural,zhao2020bridging,li2021neural}.

\begin{figure*}
    \centering
    \includegraphics[width=0.98\textwidth]{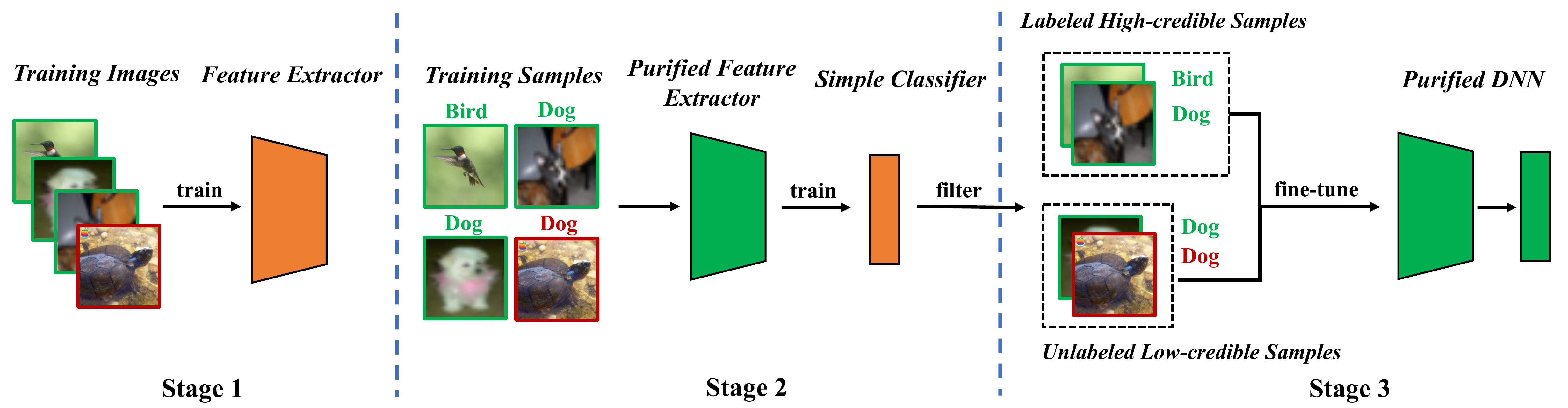}
    \vspace{-0.5em}
    \caption{The main pipeline of our defense. In the first stage, we train the whole DNN model via self-supervised learning based on label-removed training samples. In the second stage, we freeze the learned feature extractor and adopt all training samples to train the remaining fully connected layers via supervised learning. After that, we filter high-credible samples based on the training loss. In the third stage, we adopt high-credible samples as labeled samples and remove the labels of all low-credible samples to fine-tune the whole model via semi-supervised learning. }
    \vspace{-0.3em}
    \label{fig:overview}
\end{figure*}

\noindent \textbf{Defender's Goals.} 
The defender's goals are to prevent the trained DNN model from predicting poisoned samples as the target label and to preserve the high accuracy on benign samples.

\subsection{Overview of the Defense Pipeline}
In this section, we describe the general pipeline of our defense. As shown in Figure \ref{fig:overview}, it consists of three main stages, including \textbf{(1)} learning a  purified feature extractor via self-supervised learning, \textbf{(2)} filtering high-credible samples via label-noise learning, and \textbf{(3)} semi-supervised fine-tuning.

Specifically, in the first stage, we remove the label of all training samples to form the unlabelled dataset, based on which to train the feature extractor via self-supervised learning. In the second stage, we freeze the learned feature extractor and adopt all training samples to train the remaining fully connected layers via supervised learning. We then filter $\alpha\%$ high-credible samples based on the training loss. The smaller the loss, the more credible the sample. After the second stage, the training set will be separated into two disjoint parts, including high-credible samples and low-credible samples. We use high-credible samples as labeled samples and remove the label of all low-credible samples to fine-tune the whole model via semi-supervised learning. More detailed information about each stage of our method will be further illustrated in following sections.

\subsection{Learning Purified Feature Extractor via Self-supervised Learning}
\label{sec:purified}

Let $\mathcal{D}_t$ denotes the training set and $f_{\bm{w}}: \mathcal{X} \rightarrow [0,1]^K$ indicates the DNN with parameter $\bm{w}=[\bm{w}_c, \bm{w}_f]$, where $\bm{w}_c$ and $\bm{w}_f$ indicates the parameters of the backbone and the fully connected layer, respectively. In this stage, we optimize $\bm{w}_c$ based on the unlabeled version of $\mathcal{D}_t$ via self-supervised learning, as follows:
\begin{equation}\label{eq:w_c}
    \bm{w}_c^{*} = \arg \min_{\bm{w}_c} \sum_{(\bm{x},y) \in \mathcal{D}_t} \mathcal{L}_1(\bm{x};\bm{w}_c),
\end{equation}
where $\mathcal{L}_1(\cdot)$ indicates the self-supervised loss ($e.g.$, NT-Xent in SimCLR \citep{chen2020simple}). 
Through the self-supervised learning, the learned feature extractor ($i.e.$, backbone) will be purified even if the training set contains poisoned samples, as illustrated in Section \ref{sec:motivation}.

\subsection{Filtering High-credible Samples via Label-noise Learning}
\label{sec:filtering}
Once $\bm{w}_c^{*}$ is obtained, the user can freeze it and adopt $\mathcal{D}_t$ to further optimize remaining $\bm{w}_f$, $i.e.$,  
\begin{equation}\label{eq:w_f}
    \bm{w}_f^{*} = \arg \min_{\bm{w}_f} \sum_{(\bm{x},y) \in \mathcal{D}_t} \mathcal{L}_2\left(f_{[\bm{w}_c^{*},\bm{w}_f]}(\bm{x}), y \right),
\end{equation}
where $\mathcal{L}_2(\cdot)$ indicates the supervised loss ($e.g.$, cross entropy).

\begin{figure*}[ht]
\centering
\subfigure[Cross-Entropy]{
\centering
\includegraphics[width=0.475\textwidth]{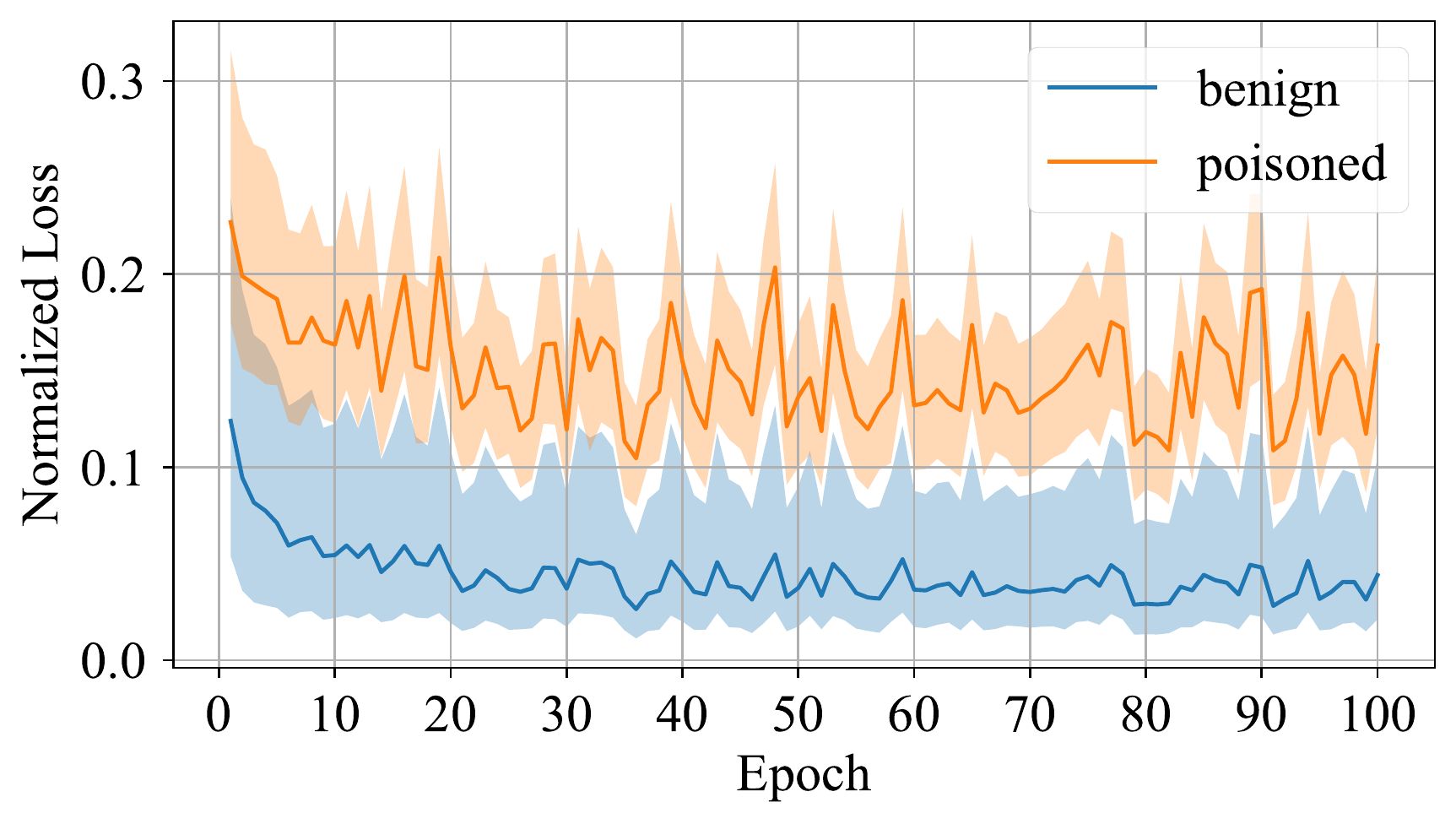}}
\hspace{0.3em}
\subfigure[Symmetric Cross-Entropy]{
\centering
\includegraphics[width=0.475\textwidth]{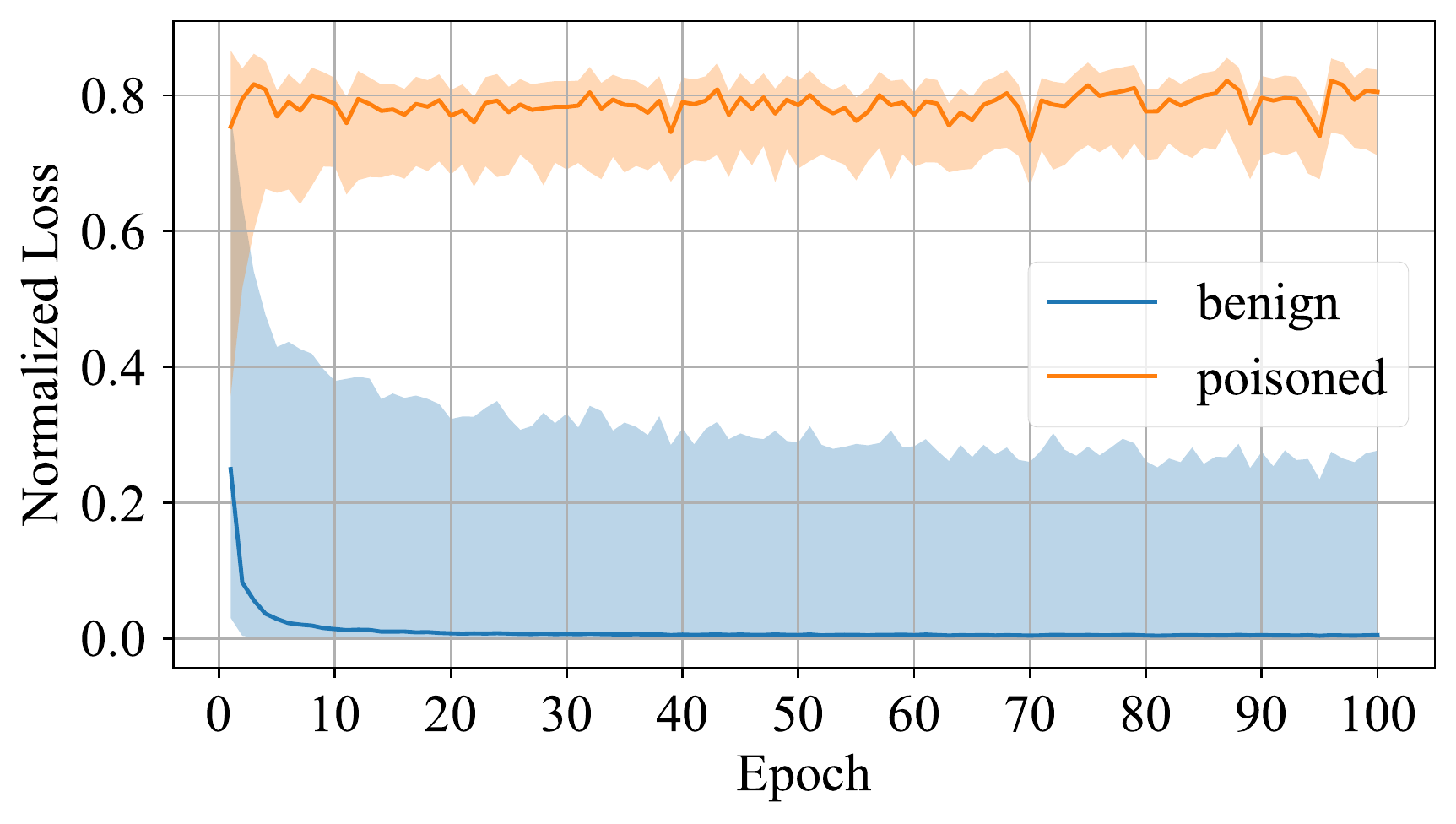}}
\vspace{-0.7em}
\caption{Loss values of models under BadNets attack with 20\% poisoning rate trained on CIFAR-10 dataset with the symmetric cross-entropy (SCE) and cross-entropy (CE) in the second stage. All loss values are normalized to $[0, 1]$. As shown in the figure, adopting SCE can significantly increase the loss differences between poisoned samples and benign ones compared with the CE. }
\label{fig:sce}
\end{figure*}

After the decoupling-based training process (\ref{eq:w_c})-(\ref{eq:w_f}), even if the model is (partly) trained on the poisoned dataset, the hidden backdoor cannot be created since the feature extractor is purified. However, this simple strategy suffers from two main problems. Firstly, compared with the one trained via supervised learning, the accuracy of predicting benign samples will have a certain decrease, since the learned feature extractor is frozen in the second stage. 
Secondly, poisoned samples will serve as `outliers' to further hinder the learning of the second stage when poison-label attacks appear, since those samples lie close to samples with its ground-truth label instead of the target label in the hidden feature space generated by the learned purified feature extractor. These two problems indicate that \emph{we should remove poisoned samples and retrain or fine-tune the whole model}.

Specifically, we select \emph{high-credible samples} $\mathcal{D}_h$ based on the loss $\mathcal{L}_2(\cdot; [\bm{w}_c^{*},\bm{w}_f^{*}])$. The high-credible samples are defined as the $\alpha\%$ training samples with the smallest loss, where $\alpha \in [0, 100]$ is a hyper-parameter. In particular, we adopt the symmetric cross-entropy (SCE) \citep{wang2019symmetric} as $\mathcal{L}_2(\cdot)$, inspired by the label-noise learning. As shown in Figure \ref{fig:sce}, compared with the CE loss, the SCE can significantly increase the differences between poisoned samples and benign ones, which further reduces the possibility that high-credible dataset $\mathcal{D}_h$ still contains poisoned samples.

Note that \emph{we do not intend to accurately separate poisoned samples and benign samples}. We only want to ensure that the obtained $\mathcal{D}_h$ contains as few poisoned samples as possible.

\subsection{Semi-supervised Fine-tuning}
\label{sec:semi}
After the second stage, the third-party training set $\mathcal{D}_t$ will be separated into two disjoint parts, including the high-credible dataset $\mathcal{D}_h$ and the low-credible dataset $\mathcal{D}_l \triangleq \mathcal{D}_t \backslash \mathcal{D}_h$. Let $\hat{\mathcal{D}}_l \triangleq \{\bm{x}|(\bm{x}, y) \in \mathcal{D}_l\}$ indicates the unlabeled version of low-credible dataset $\mathcal{D}_l$. We fine-tune the whole trained model $f_{[\bm{w}_c^{*},\bm{w}_f^{*}]}(\cdot)$ with semi-supervised learning as follows:
\begin{equation}
    \min_{\bm{w}} \mathcal{L}_3 (\mathcal{D}_h,\hat{\mathcal{D}}_l;\bm{w} ),
\end{equation}
where 
$\mathcal{L}_3(\cdot)$ denotes the semi-supervised loss ($e.g.$, the loss in MixMatch \citep{berthelot2019mixmatch}). 

This process can prevent the side-effects of poisoned samples while utilizing their contained useful information, and encourage the compatibility between the feature extractor and the simple classifier via learning them jointly instead of separately. Please refer to Section \ref{sec:ablation_study} for more results.

\section{Experiments}

\subsection{Experimental Settings}
\noindent \textbf{Datasets and DNNs.}
We evaluate all defenses on two classical benchmark datasets, including CIFAR-10 \citep{krizhevsky2009learning} and (a subset of) ImageNet \citep{deng2009imagenet}. We adopt the ResNet-18 \citep{he2016deep} for these tasks. More detailed settings are presented in Appendix \ref{sec:DD_set}. Besides, we also provide the results on (a subset of) VGGFace2 \citep{cao2018vggface2} in Appendix \ref{sec:results_vggface2}.

\begin{figure*}
    \centering
    \includegraphics[width=0.98\textwidth]{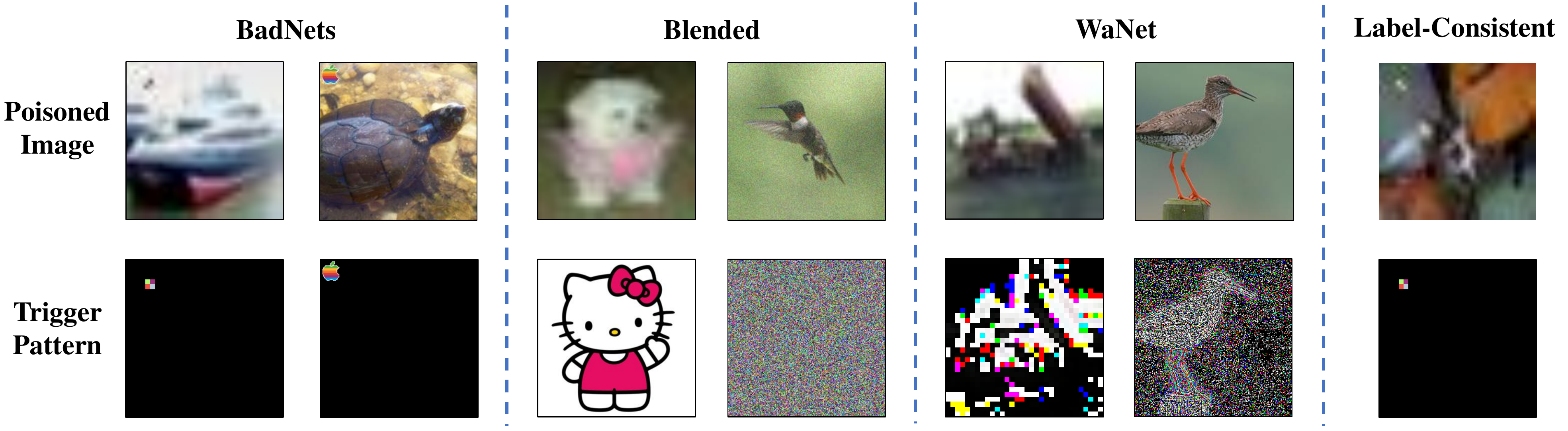}
    \vspace{-0.7em}
    \caption{The illustration of poisoned samples generated by different attacks. }
    \label{fig:poisoned_samples}
\end{figure*}

\noindent \textbf{Attack Baselines.} 
We examine all defense approaches in defending against four representative attacks. Specifically, we select the BadNets \citep{gu2017badnets}, the backdoor attack with blended strategy (dubbed `Blended') \citep{chen2017targeted}, WaNet \citep{nguyen2021wanet}, and label-consistent attack with adversarial perturbations (dubbed `Label-Consistent') \citep{turner2019label} for the evaluation. They are the representative of patch-based visible and invisible poison-label attacks, non-patch-based poison-label attacks, and clean-label attacks, respectively.

\noindent \textbf{Defense Baselines.} 
We compared our DBD with two defenses having the same defender's capacities, including the DPSGD \citep{du2019robust} and ShrinkPad \citep{li2021backdoor2}. 
We also compare with other two approaches with an additional requirement ($i.e.$, having a local benign dataset), including the neural cleanse with unlearning strategy (dubbed `NC') \citep{wang2019neural}, and neural attention distillation (dubbed `NAD') \citep{li2021neural}. 
They are the representative of poison suppression based defenses, preprocessing based defenses, trigger synthesis based defenses, and model reconstruction based defenses, respectively. We also provide results of DNNs trained without any defense (dubbed `No Defense') as another important baseline for reference.

\noindent \textbf{Attack Setups.} 
We use a $2\times 2$ square as the trigger pattern on CIFAR-10 dataset and the $32 \times 32$ Apple logo on ImageNet dataset for the BadNets, as suggested in \citep{gu2017badnets,wang2019neural}. For Blended, we adopt the `Hello Kitty' pattern on CIFAR-10 and the random noise pattern on ImageNet, based on the suggestions in \citep{chen2017targeted}, and set the blended ratio $\lambda = 0.1$ on all datasets. The trigger pattern adopted in label-consistent attack is the same as the one used in BadNets. For WaNet, we adopt its default settings on CIFAR-10 dataset. However, on ImageNet dataset, we use different settings optimized by grid-search since the original ones fail. An example of poisoned samples generated by different attacks is shown in Figure \ref{fig:poisoned_samples}. Besides, we set the poisoning rate $\gamma_1=2.5\%$ for label-consistent attack (25\% of training samples with the target label) and $\gamma_2= 5\%$ for three other attacks. More details are shown in Appendix \ref{sec:attack_set}.

\noindent \textbf{Defense Setups.} 
For our DBD, we adopt SimCLR \citep{chen2020simple} as the self-supervised method and MixMatch \citep{berthelot2019mixmatch} as the semi-supervised method. More details about SimCLR and MixMatch are in Appendix \ref{sec:more_loss}. The filtering rate $\alpha$ is the only key hyper-parameter in DBD, which is set to 50\% in all cases. We set the shrinking rate to 10\% for the ShrinkPad on all datasets, as suggested in \citep{li2021backdoor2,zeng2021deepsweep}. In particular, DPSGD and NAD are sensitive to their hyper-parameters. We \emph{report their best results in each case} based on the grid-search (as shown in Appendix \ref{sec:search_best}). Besides, we split a 5\% random subset of the benign training set as the local benign dataset for NC and NAD. More implementation details are provided in Appendix \ref{sec:defense_set}.

\noindent \textbf{Evaluation Metrics.} 
We adopt the \emph{attack success rate} (ASR) and \emph{benign accuracy} (BA) to measure the effectiveness of all methods\footnote{Among all defense methods, the one with the best performance is indicated in boldface and the value with underline denotes the second-best result.}. Specifically, let $\mathcal{D}_{test}$ indicates the (benign) testing set and $C_{\bm{w}}: \mathcal{X} \rightarrow \mathcal{Y}$ denotes the trained classifier, we have $ASR \triangleq \Pr_{(\bm{x},y) \in \mathcal{D}_{test}}\{C_{\bm{w}}(G(\bm{x}))=y_t|y \neq y_t\}$ and $BA \triangleq \Pr_{(\bm{x},y) \in \mathcal{D}_{test}}\{C_{\bm{w}}(\bm{x})=y\}$, where $y_t$ is the target label and $G(\cdot)$ is the poisoned image generator. In particular, \emph{the lower the ASR and the higher the BA, the better the defense}.

\subsection{Main Results}
\label{sec:main}

\noindent \textbf{Comparing DBD with Defenses having the Same Requirements.}
As shown in Table \ref{tab:poison}-\ref{tab:clean}, DBD is significantly better than defenses having the same requirements ($i.e.$, DPSGD and ShrinkPad) in defending against all attacks. For example, the benign accuracy of DBD is 20\% over while the attack success rate is 5\% less than that of DPSGD in all cases. Specifically, the attack success rate of models with DBD is less than 2\% in all cases (mostly $< 0.5\%$), which verifies that our method can successfully prevent the creation of hidden backdoors. Moreover, the decreases of benign accuracy are less than $2\%$ when defending against poison-label attacks, compared with models trained without any defense. Our method is even better on relatively larger dataset where all baseline methods become less effective. These results verify the effectiveness of our method.

\begin{table*}[ht]
\centering
\vspace{-2.8em}
\caption{The effectiveness (\%) of defending against three attacks. Note that NC and NAD need an additional local benign dataset, which is not required in DPSGD, ShrinkPad, and our method.}
\scalebox{0.83}{
\begin{tabular}{c|cccccc|cccccc}
\toprule
Dataset $\rightarrow$ & \multicolumn{6}{c|}{CIFAR-10} & \multicolumn{6}{c}{ImageNet} \\ \hline
Attack $\rightarrow$ & \multicolumn{2}{c}{BadNets} & \multicolumn{2}{c}{Blended} &
\multicolumn{2}{c|}{WaNet} & \multicolumn{2}{c}{BadNets} &  \multicolumn{2}{c}{Blended} & \multicolumn{2}{c}{WaNet} \\ \hline
Defense $\downarrow$ & BA & ASR & BA & ASR & BA & ASR & BA & ASR & BA & ASR & BA & ASR \\ \hline
No Defense & 94.92 & 100 & 94.14 & 98.25 & 94.29 & 98.64 & 80.23 & 90.49 & 82.07 & 95.57 & 80.98 & 96.22 \\ \hline
NC & \textbf{94.25} & 2.33 & \textbf{93.01} & 5.70 & \textbf{92.44} & 98.52 & \underline{80.67} & 32.22 & \textbf{82.07} & 27.33 & \underline{80.32} & 99.32 \\
NAD & 90.47 & \underline{1.20} & 89.72 & \textbf{1.51} & \underline{91.83} & \underline{9.49} & 73.49 & \underline{5.81} & 70.88 & \underline{8.15} & 75.27 & 31.43 \\ \hline
DPSGD & 44.24 & 14.56 & 53.05 & 20.68 & 52.13 & 39.78 & 51.43 & 25.40 & 55.12 & 84.17 & 20.72 & 54.12  \\
ShrinkPad & 90.32 & 10.34 & 79.67 & 55.71 & 74.76 & 85.59 & 66.08 & 45.61 & 64.47 & 4.20 & 63.74 & \underline{3.58} \\
DBD (ours) & \underline{92.41} & \textbf{0.96} & \underline{92.18} & \underline{1.73} & 91.20 & \textbf{0.39} & \textbf{80.99} & \textbf{0.26} & \underline{81.63} & \textbf{0.16} & \textbf{82.05} & \textbf{0.33} \\ \bottomrule
\end{tabular}
}
\vspace{-1.5em}
\label{tab:poison}
\end{table*}

\begin{table*}[ht]
\centering
\caption{The effectiveness (\%) of defending against label-consistent attack on CIFAR-10 dataset. 
}
\scalebox{0.9}{
        \begin{tabular}{c|c|c|cc|ccc}
            \toprule
            Attack$\downarrow$, Defense$\rightarrow$                               & Metric$\downarrow$ & Benign & NC               & NAD   & DPSGD & ShrinkPad & DBD (ours)        \\ \hline
            \multirow{2}{*}{\tabincell{c}{Label-Consistent\\($\epsilon=16$)}} & BA                                         & 94.90  & \textbf{94.94}   & 93.04 & 44.28 & \underline{90.67}     & 89.67             \\
                                                              & ASR                                        & 99.33  & \underline{1.31} & 3.95  & 7.97  & 9.60      & \textbf{0.01}     \\ \hline
            \multirow{2}{*}{\tabincell{c}{Label-Consistent\\($\epsilon=32$)}} & BA                                         & 94.82  & \textbf{94.57}   & 90.98 & 69.79 & 90.83     & \underline{91.45} \\
                                                              & ASR                                        & 99.19  & \underline{1.36} & 6.54  & 9.06  & 13.04     & \textbf{0.34}     \\ \bottomrule
        \end{tabular}
    }
\label{tab:clean}
\vspace{-0.9em}
\end{table*}

\noindent \textbf{Comparing DBD with Defenses having Extra Requirements.}
We also compare our defense with two other methods ($i.e.$, NC and NAD), which have an additional requirement that defenders have a benign local dataset. As shown in Table \ref{tab:poison}-\ref{tab:clean}, NC and NAD are better than DPSGD and ShrinkPad, as we expected, since they adopt additional information from the benign local dataset. In particular, although NAD and NC use additional information, our method is still better than them, even when their performances are tuned to the best while our method only uses the default settings. Specifically, the BA of NC is on par with that of our method. However, it is with the sacrifice of ASR. Especially on ImageNet dataset, NC has limited effects in reducing ASR. In contrast, our method reaches the smallest ASR while its BA is either the highest or the second-highest in almost all cases. These results verify the effectiveness of our method again.

\begin{table*}[!t]
   \centering
   \vspace{-0.5em}
   \caption{The ablation study of our proposed method.}
   \scalebox{0.95}{
      \begin{tabular}{c|cccc|cc|cc}
         \toprule
         Attack $\rightarrow$                       & \multicolumn{2}{c}{BadNets} & \multicolumn{2}{c|}{Blended} & \multicolumn{2}{c|}{Label-Consistent} & \multicolumn{2}{c}{WaNet}\\ \hline
         Defense $\downarrow$, Metric $\rightarrow$ & BA                          & ASR                          & BA                                    & ASR                                  & BA                                    & ASR   & BA    & ASR\\ \hline
         No Defense                                 & 94.92                       & 100                          & 94.14                                 & 98.25                                & 94.82                                 & 99.19 & 94.29 & 98.64\\ \hline
         DBD without SS   & \textbf{93.66}                       & 100                          & \textbf{93.47}                                 & 99.93                                & 90.70                                 & 98.50  & 81.91 & 98.40\\ \hline
         SS with CE                                 & 82.25                       & 5.20                         & 81.85                                 & 12.19                                & 82.08                                 & 5.87  & 80.29 & 9.48\\
         SS with SCE                                & 82.34                       & 5.12                         & 82.30                                 & 6.24                                 & 81.81                                 & 5.43  & 81.15 & 7.08\\ \hline
         SS with SCE + Tuning                       & 78.94                       & \underline{4.02}                         & 78.78                                 & \underline{3.70}                                 & 77.44                                 & \underline{1.44}  & 78.51 & \underline{5.87}\\ \hline
         DBD (ours)                                 & \underline{92.41}                       & \textbf{0.96}                         & \underline{92.18}                                 & \textbf{1.73}                                 & \textbf{91.45}                                 & \textbf{0.34}  & \textbf{91.20} & \textbf{0.39}
         \\ \bottomrule
      \end{tabular}
   }
   \begin{tablenotes}
      \footnotesize
      \item[1] $^{1}$DBD without SS: use standard supervised training to obtain the feature extractor adopted in DBD.
      \item[2] $^{2}$SS with CE: freeze the learned backbone and train remaining FC layers on all samples with CE loss.
      \item[3] $^{3}$SS with SCE: freeze the learned backbone and train remaining FC layers on all samples with SCE loss.
      \item[4] $^{4}$SS with SCE + Tuning: fine-tune FC layers of the model in `SS with SCE' on high-credible samples.
   \end{tablenotes}
   \label{tab:ablation}
   \vspace{-1em}
\end{table*}

\noindent \textbf{Results. }
As shown in Figure \ref{fig:gamma}, our method can still prevent the creation of hidden backdoors even when the poisoning rate reaches 20\%. Besides, DBD also maintains high benign accuracy. In other words, our method is effective in defending attacks with different strengths. 

\subsection{Ablation Study}
\label{sec:ablation_study}

There are four key strategies in DBD, including \textbf{(1)} obtaining purified feature extractor, \textbf{(2)} using SCE instead of CE in the second stage, \textbf{(3)} reducing side-effects of low-credible samples, and \textbf{(4)} fine-tuning the whole model via semi-supervised learning. Here we verify their effectiveness.

\noindent \textbf{Settings. }
We compare the proposed DBD with its four variants, including \textbf{(1)} DBD without SS, \textbf{(2)} SS with CE, \textbf{(3)} SS with SCE, and \textbf{(4)} SS with SCE + Tuning, on the CIFAR-10 dataset. Specifically, in the first variant, we replace the backbone generated by self-supervised learning with the one trained in a supervised fashion and keep other parts unchanged. In the second variant, we freeze the backbone learned via self-supervised learning and train the remaining fully-connected layers with cross-entropy loss on all training samples. The third variant is similar to the second one. The only difference is that it uses symmetric cross-entropy instead of cross-entropy to train fully-connected layers. The last variant is an advanced version of the third one, which further fine-tunes fully-connected layers on high-credible samples filtered by the third variant. 

\noindent \textbf{Results. }
As shown in Table \ref{tab:ablation}, we can conclude that decoupling the original end-to-end supervised training process is effective in preventing the creation of hidden backdoors, by comparing our DBD with its first variant and the model trained without any defense. Besides, we can also verify the effectiveness of SCE loss on defending against poison-label backdoor attacks by comparing the second and third DBD variants. Moreover, the fourth DBD variant has relatively lower ASR and BA, compared with the third one. This phenomenon is due to the removal of low-credible samples. It indicates that reducing side-effects of low-credible samples while adopting their useful information is important for the defense. We can also verify that fine-tuning the whole model via semi-supervised learning is also useful by comparing the fourth variant and the proposed DBD.

\subsection{Resistance to Potential Adaptive Attacks}
\label{sec:adaptive}

In our paper, we adopted the classical defense setting that attackers have no information about the defense. Attackers may design adaptive attacks if they know the existence of our DBD. The most straightforward idea is to \emph{manipulate the self-supervised training process} so that poisoned samples are still in a new cluster after the self-supervised learning. However, \emph{attackers are not allowed to do it} based on our threat model about adopting third-party datasets. Despite this, attackers may design adaptive attacks by optimizing the trigger pattern to make poisoned samples still in a new cluster after the self-supervised learning if they can know the model structure used by defenders, as follows:

\noindent \textbf{Problem Formulation.} For a $K$-classification problem, let $\mathcal{X}'=\{\bm{x}_i\}_{i=1}^M$ indicates the benign images selected for poisoning,  $\mathcal{X}_{j}=\{\bm{x}_i\}_{i=1}^{N_j}$ denotes the benign images with ground-truth label $j$, and $g$ is a trained backbone. Given an attacker-predefined poisoned image generator $G$, the \emph{adaptive attack} aims to optimize a trigger pattern $\bm{t}$ by minimizing the distance between poisoned images while maximizing the distance between the center of poisoned images and centers of clusters of benign images with different label, $i.e.,$
\begin{equation}
\min_{\bm{t}} \frac{1}{M}\sum_{\bm{x} \in \mathcal{X}'} d\left(g(G(\bm{x};\bm{t})), \overline{g}')\right) - \frac{1}{K}\sum_{i=1}^K d\left(\overline{g}', \overline{g_{i}}\right),     
\end{equation}
    
where $\overline{g}' \triangleq \frac{1}{M}\sum_{\bm{x} \in \mathcal{X}'} g(G(\bm{x};\bm{t}))$, $\overline{g_i} \triangleq \frac{1}{N_i} \sum_{\bm{x} \in \mathcal{X}_i} g(\bm{x})$, and $d$ is a distance metric.

\noindent \textbf{Settings.}
We adopt the CIFAR-10 dataset and use the $\ell^2$ norm as the distance metric to conduct the experiment. Specifically, we assume that attackers have the entire benign dataset, based on which they can train a backbone adopted in the first stage of our DBD. We use the Adam optimizer to solve the above optimization problem for 100 epochs with a learning rate of 0.1. The trigger size is set to $32\times32$, which means the attacker can completely modify the content of poisoned samples, regardless of its original semantic information and the stealthiness of the attack. This setting is to ensure the attack ability, since clustering poisoned samples together is very difficult in self-supervised learning.

\noindent \textbf{Results.}
The adaptive attack works well when there is no defense (BA=94.96\%, ASR=99.70\%). However, this attack still fails to attack our DBD (BA=93.21\%, ASR=1.02\%). In other words, our defense is resistant to this adaptive attack. It is most probably because the trigger optimized based on the backbone is far less effective when the model is retrained since model parameters are changed due to the random initialization and the update of model weights during the training process.

\comment{
\subsection{A Brief Description of Additional Experiments in the Appendix}
Except for above experiments, we also provide additional results of defending against attacks with different hyper-parameters in Appendix \ref{sec:smaller_p}-\ref{sec:diff_trigger} and the resistance to adaptive attacks in Appendix \ref{sec:adaptive}. 
}

\vspace{-0.3em}
\section{Conclusion}
\vspace{-0.2em}

The mechanism of poisoning-based backdoor attacks is to establish a latent connection between trigger patterns and the target label during the training process. In this paper, we revealed that this connection is learned mostly due to the end-to-end supervised training paradigm. Motivated by this understanding, we proposed a decoupling-based backdoor defense, which first learns the backbone via self-supervised learning and then the remaining fully-connected layers by the classical supervised learning. We also introduced the label-noise learning method to determine high-credible and low-credible samples, based on which we fine-tuned the whole model via semi-supervised learning. Extensive experiments verify that our defense is effective on reducing backdoor threats while preserving high accuracy on predicting benign samples. 

\newpage

\section*{Acknowledgments}
Baoyuan Wu is supported in part by the National Natural Science Foundation of China under Grant 62076213, the University Development Fund of the Chinese University of Hong Kong, Shenzhen under Grant 01001810, and the Special Project Fund of Shenzhen Research Institute of Big Data under Grant T00120210003. 
Zhan Qin is supported in part by the National Natural Science Foundation of China under Grant U20A20178, the National Key Research and Development Program of China under Grant 2020AAA0107705, and the Research Laboratory for Data Security and Privacy, Zhejiang University-Ant Financial Fintech Center.
Kui Ren is supported by the National Key Research and Development Program of China under Grant 2020AAA0107705.


\section*{Ethics Statement}
DNNs are widely adopted in many mission-critical areas ($e.g.$, face recognition) and therefore their security is of great significance. The vulnerability of DNNs to backdoor attacks raises serious concerns about using third-party training resources. In this paper, we propose a general training pipeline to obtain backdoor-free DNNs, even if the training dataset contains poisoned samples. This work has no ethical issues in general since our method is purely defensive and does not reveal any new vulnerabilities of DNNs. However, we need to mention that our defense can be adopted only when training with untrusted samples, and backdoor attacks could happen in other scenarios. People should not be too optimistic about eliminating backdoor threats.

\section*{Reproducibility Statement}
The detailed descriptions of datasets, models, and training settings are in Appendix \ref{sec:motivation_set}-\ref{sec:search_best}. We also describe the computational facilities and cost in Appendix \ref{sec:train_f}-\ref{sec: computational cost}. Codes of our DBD are also open-sourced.

\bibliography{iclr2022_conference}
\bibliographystyle{iclr2022_conference}

\newpage

\appendix

\begin{table*}[th]
      \centering
      \vspace{-1em}
      \caption{Statistics of datasets and DNNs adopted in our main experiments.}
      \scalebox{0.9}{
      \begin{tabular}{cccccc}
            \toprule
            Dataset  & Input Size      & \# Classes & \# Training Images & \# Test Images & DNN model    \\ \midrule
            CIFAR-10 & $3\times 32\times 32$   & 10         & 50,000              & 10,000          & ResNet-18    \\ \hline
            ImageNet & $3\times 224\times 224$ & 30         & 38,939              & 1,500           & ResNet-18    \\ \bottomrule
      \end{tabular}
      }
      \label{tab:stat}
      \vspace{-1em}
\end{table*}

\section{Detailed Settings for Revisiting Backdoor Attacks}
\label{sec:motivation_set}

\noindent \textbf{Attack Setups.}
We conduct the BadNets \citep{gu2017badnets} and label-consistent attack \citep{turner2019label} with the target label $y_t=3$ on the CIFAR-10 dataset \citep{krizhevsky2009learning}. The trigger patterns are the same as those presented in Section \ref{sec:main}. In particular, we implement the label-consistent attack with adversarial perturbations, as suggested in its original paper \citep{turner2019label}. Specifically, we used the projected gradient descent (PGD) \citep{madry2018towards} to generate adversarial perturbations within the $\ell^{\infty}$-ball where the maximum perturbation size $\epsilon=16$.

\noindent \textbf{Training Setups.}
We conduct supervised learning on the poisoned datasets with the standard training process and the self-supervised learning on the unlabelled poisoned datasets with the SimCLR \citep{chen2020simple}. The supervised training is conducted based on the open-source code\footnote{\url{https://github.com/kuangliu/pytorch-cifar}}. Specifically, we use the SGD optimizer with momentum 0.9, weight decay of $5 \times 10^{-4}$, and an initial learning rate of 0.1. The batch size is set to 128 and we train the ResNet-18 model 200 epochs. The learning rate is decreased by a factor of 10 at epoch 100 and 150, respectively. Besides, we add triggers before performing the data augmentation ($e.g.$, random crop and horizontal flipping). For the self-supervised training, we use the stochastic gradient descent (SGD) optimizer with a momentum of 0.9, an initial learning rate of 0.4, and a weight decay factor of $5\times 10^{-4}$. We use a batch size of 512, and train the backbone for 1,000 epochs. We decay the learning rate with the cosine decay schedule \citep{loshchilov2016sgdr} without a restart. Besides, we also adopt strong data augmentation techniques, including random crop and resize (with random flip), color distortions, and Gaussian blur, as suggested in \citep{chen2020simple}. All models are trained until converge.

\noindent \textbf{t-SNE Visualization Settings.}
We treat the output of the last residual unit as the feature representation and use the tsne-cuda library \citep{chan2019gpu} to get the feature embedding of all samples. To have a better visualization, we adopt all poisoned samples and randomly select 10\% benign samples for visualizing models under the supervised learning, and adopt 30\% poisoned samples and 10\% benign samples for those under the self-supervised learning.

\section{Detailed Settings for Main Experiments}
\subsection{More Details about Datasets and DNNs}
\label{sec:DD_set}

Due to the limitations of computational resources and time, we adopt a subset randomly selected from the original ImageNet. More detailed information about the datasets and DNNs adopted in the main experiments of our paper is presented in Table \ref{tab:stat}.

\subsection{More Details about Attack Settings}
\label{sec:attack_set}

\noindent \textbf{Attack Setups.}
We conduct the BadNets \citep{gu2017badnets}, blended attack (dubbed `Blended') \citep{chen2017targeted}, label-consistent attack (dubbed `Label-Consistent') \citep{turner2019label}, and WaNet \citep{nguyen2021wanet} with the target label $y_t=3$ on all datasets. The trigger patterns are the same as those presented in Section \ref{sec:main}. In particular, we set the blended ratio $\lambda = 0.1$ for the blended attack on all datasets and examine label-consistent attack with the maximum perturbation size $\epsilon \in \{16, 32\}$. Besides, WaNet assumed that attackers can fully control the whole training process in its original paper. However, we found that WaNet only modified training data while other training components ($e.g.$, training loss, training schedule, and model structure) are the same as those used in the standard training process. As such, we re-implement its code in the poisoning-based attack scenario based on its official code\footnote{\url{https://github.com/VinAIResearch/Warping-based\_Backdoor\_Attack-release}}. Specifically, following the settings in its original paper, we set the noise rate $\rho_n=0.2$, control grid size $k=4$, and warping strength $s=0.5$ on the CIFAR-10 dataset. However, we found that the default $k$ and $s$ are too small to make the attack works on the ImageNet dataset (as shown in Table \ref{tab:k_effect}-\ref{tab:s_effect}). Besides, the `noise mode' also significantly reduces the attack effectiveness (as shown in Table \ref{tab:noise_effect}). As such, we set $k=224$ and $s=1$ and train models without the noise mode on the ImageNet dataset.

\begin{table*}[!t]
   \centering
   \caption{The Effectiveness of WaNet with different kernel size $k$ when the noise rate is set to 0 and strength $s=1$ on the ImageNet dataset. }
   \begin{tabular}{c|cccc}
      \toprule
      $k$ $\rightarrow$ & 4     & 32    & 128   & 224   \\ \hline
      BA              & 78.31 & 77.64 & \textbf{81.29} & 80.98 \\ \hline
      ASR             & 6.87  & 14.98 & 91.69 & \textbf{96.22} \\ \bottomrule
   \end{tabular}
   \label{tab:k_effect}
   \vspace{-1em}
\end{table*}

\begin{table*}[!t]
   \centering
   \caption{The Effectiveness of WaNet with different strength $s$ when the noise rate is set to 0 and strength kernel size $k=224$ on the ImageNet dataset. }
   \begin{tabular}{c|cccc}
      \toprule
      $s$ $\rightarrow$ & 0.4 & 0.6 & 0.8 & 1 \\ \hline
      BA  & 81.01 & \textbf{81.45} & 81.32 & 80.98 \\ \hline
      ASR & 80.06 & 92.22 & 94.23 & \textbf{96.22} \\ \bottomrule
   \end{tabular}
   \label{tab:s_effect}
   \vspace{-1em}
\end{table*}

\begin{table*}[!t]
   \centering
   \caption{The Effectiveness of WaNet with and without the noise mode when the kernel size $k=224$ and strength $s=1$ on the ImageNet dataset. }
   \begin{tabular}{c|cc}
      \toprule
          & w/ noise mode & w/o noise mode \\ \hline
      BA  & 79.37        & \textbf{80.98}          \\ \hline
      ASR & 31.72        & \textbf{96.22}          \\ \bottomrule
   \end{tabular}
   \label{tab:noise_effect}
   \vspace{-0.8em}
\end{table*}

\noindent \textbf{Training Setups.}
On the CIFAR-10 dataset \citep{krizhevsky2009learning}, the settings are the same as those described in Section \ref{sec:motivation_set}; On the ImageNet dataset \citep{deng2009imagenet}, we conduct experiments based on the open-source code\footnote{\url{https://github.com/pytorch/examples/tree/master/imagenet}}. Specifically, we use the SGD optimizer with momentum 0.9, weight decay of $10^{-4}$, and an initial learning rate of 0.1. The batch size is set to 256 and we train the ResNet-18 model 90 epochs. The learning rate is decreased by a factor of 10 at epoch 30 and 60, respectively. Besides, since the raw images in the ImageNet dataset are of different sizes, we resize them to $3 \times 224\times 224$ before adding triggers.

\subsection{More Details about Defense Settings}
\label{sec:defense_set}

\noindent \textbf{Settings for NC.}
We conduct reverse engineering and anomaly detection based on its open-source code\footnote{\url{https://github.com/bolunwang/backdoor}}. We implement the `unlearning' method to patch attacked models, as suggested in its paper \citep{wang2019neural}. We randomly select 5\% benign training samples as the local benign dataset, which is used in the `unlearning' process. Unless otherwise specified, other settings are the same as those used in \citep{wang2019neural}.

\noindent \textbf{Settings for NAD.}
We implement this method based on its open-source code\footnote{\url{https://github.com/bboylyg/NAD}}. The origin NAD only conducted experiments on the WideResNet model. In our paper, we calculate the NAD loss over the last residual group for the ResNet-18. The local benign dataset is the same as the one adopted in NC, which is used in the fine-tuning and distillation process of NAD. Unless otherwise specified, other settings are the same as those used in \citep{li2021neural}.

\noindent \textbf{Settings for DPSGD.}
The original DPSGD was conducted on the MNIST dataset implemented by the TensorFlow Framework. In this paper, we re-implement it based on the differentially private SGD method provided by the Opacus\footnote{\url{https://github.com/pytorch/opacus}}. Specifically, we replace the original SGD optimizer with the differentially private one, as suggested in \citep{du2019robust}. There are two important hyper-parameters in DPSGD, including noise scales $\sigma$ and the clipping bound $C$. In the experiments, we set $C=1$ and select the best $\sigma$ by the grid-search.

\begin{figure*}
    \centering
    \vspace{-1em}
    \includegraphics[width=0.5\textwidth]{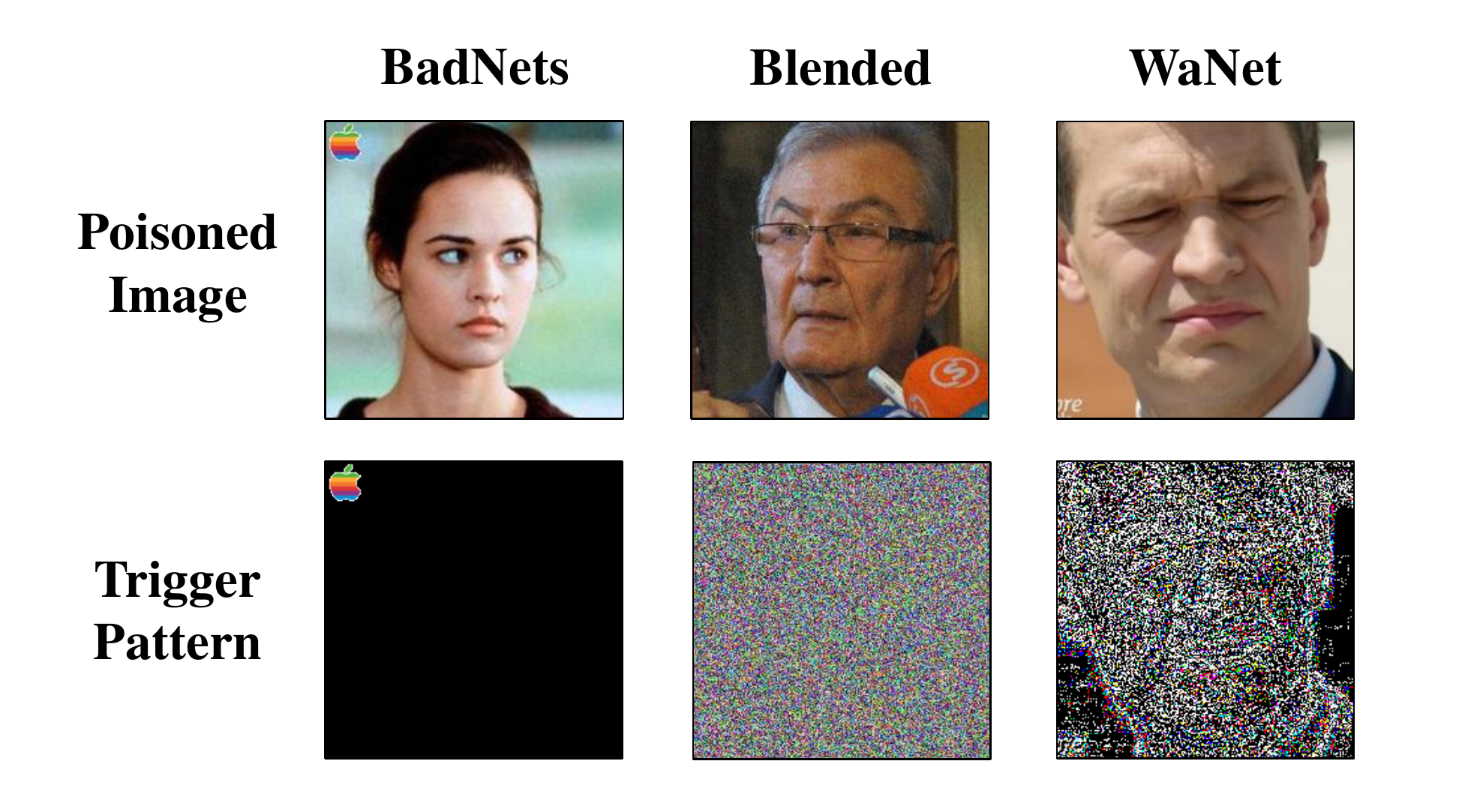}
    \vspace{-1.5em}
    \caption{An example of poisoned samples generated by different attacks on VGGFace2 dataset. }
    \label{fig:poisoned_samples_vgg}
    \vspace{-0.5em}
\end{figure*}

\begin{table*}[!t]
      \centering
      \vspace{-1em}
      \caption{Statistics of the VGGFace2 dataset and model structure adopted in our experiments.}
      \scalebox{0.9}{
      \begin{tabular}{cccccc}
            \toprule
            Dataset  & Input Size      & \# Classes & \# Training Images & \# Test Images & DNN model    \\ \midrule
            VGGFace2 & $3 \times 224\times 224$ & 30         & 9,051               & 2,263           & DenseNet-121 \\ \bottomrule
      \end{tabular}
      }
      \label{tab:stat_vgg}
\end{table*}

\begin{table*}[!ht]
\centering
\vspace{-1em}
\caption{Defending against attacks on VGGFace2 dataset. Note that NC and NAD need an additional local benign dataset, which is not required in DPSGD, ShrinkPad, and our method.}
\scalebox{0.95}{
\begin{tabular}{c|cccccc}
\toprule
Attack $\rightarrow$ & \multicolumn{2}{c}{BadNets} & \multicolumn{2}{c}{Blended} & \multicolumn{2}{c}{WaNet} \\ \hline
Defense $\downarrow$, Metric $\rightarrow$ & BA & ASR & BA & ASR & BA & ASR \\ \hline
No Defense & 91.52 & 88.14 & 93.35 & 98.94 & 91.41 & 98.75\\ \hline
NC & \textbf{92.16} & 18.82 & \textbf{93.43} & 86.76 & \textbf{91.32} & 95.09 \\
NAD & 87.29 & \underline{9.49} & 85.27 & \underline{10.25} & 80.69 & \underline{4.41} \\ \hline
DPSGD & 60.83 & 6.61 & 81.90 & 93.63 & 54.04 & 30.85 \\
ShrinkPad & 77.64 & 61.50 & 79.18 & 75.61 & 78.62 & 30.13 \\
DBD (ours) & \underline{89.74} & \textbf{0.17} & \underline{93.29} & \textbf{0.09} & \underline{89.32} & \textbf{0} \\ \bottomrule
\end{tabular}
}
\vspace{-1em}
\label{tab:main_vgg}
\end{table*}

\noindent \textbf{Settings for ShrinkPad.}
We set the shrinking rate to 10\% on all datasets, as suggested in \citep{li2021backdoor2,zeng2021deepsweep}. Following their settings, we pad 0-pixels at the bottom right of the shrunk image to expand it to its original size.

\noindent \textbf{Settings for our Defense.}
\label{sec:DBD_set}
In this first stage, We adopt SimCLR \citep{chen2020simple} to perform self-supervised learning. We train backbones 100 instead of 1,000 epochs to reduce computational costs while preserving effectiveness. Other settings are the same as those described in Section \ref{sec:motivation_set}. We use the same settings across all datasets, models, and attacks; In the second stage, we use the Adam optimizer with a learning rate of 0.002 and set the batch size to 128. We train the fully connected layers 10 epochs with the SCE loss \citep{wang2019symmetric}. Two hyper-parameters involved in the SCE ($i.e.$, $\alpha$ and $\beta$ in the original paper) are set to 0.1 and 1, respectively.
After that, we filter 50\% high-credible samples. We use the same settings across all datasets, models, and attacks; In the third stage, we adopt the MixMatch \citep{berthelot2019mixmatch} for semi-supervised fine-tuning with settings suggested in its original paper. Specifically, we use the Adam optimizer with a learning rate of 0.002, the batch size of 64, and finetune the model 190 epochs on the CIFAR-10 and 80 epochs on the ImageNet dataset, respectively. We set the temperature $T=0.5$ and the weight of unsupervised loss $\lambda_u=15$ on the CIFAR-10 and $\lambda_u=6$ on the ImageNet dataset, respectively. Moreover, we re-filter high-credible samples after every epoch of the third stage based on the SCE loss.

\section{Defending against Attacks on VGGFace2 Dataset}
\label{sec:results_vggface2}

\noindent \textbf{Dataset and DNN.}
Due to the limitations of computational resources and time, we adopt a subset randomly selected from the original VGGFace2 \citep{cao2018vggface2}. More details are in Table \ref{tab:stat_vgg}.

\noindent \textbf{Settings for Attacks.}
For the training of models on the VGGFace2 dataset, the batch size is set to 32 and we conduct experiments on the DenseNet-121 model \citep{huang2017densely}. An example of poisoned samples generated by different attacks are in Figure \ref{fig:poisoned_samples_vgg}. Other settings are the same as those used on the ImageNet dataset.

\noindent \textbf{Settings for Defenses.}
For NAD, we calculate the NAD loss over the second to last layer for the DenseNet-121. Other settings are the same as those used on the ImageNet dataset.

\begin{table*}[ht]
      \centering
      \vspace{-1.5em}
      \caption{Results of DPSGD against the BadNets and blended attack with different noise scale $\sigma$. }
      \scalebox{0.83}{
      \begin{tabular}{@{}ccccccccccccc@{}}
            \toprule
            Dataset $\rightarrow$                        & \multicolumn{4}{c}{CIFAR-10} & \multicolumn{4}{c}{ImageNet} & \multicolumn{4}{c}{VGGFace2}                                                                                                                                                                                                                                                                   \\ \midrule
            Attack $\rightarrow$                         & \multicolumn{2}{c}{BadNets}  & \multicolumn{2}{c}{Blended}  & \multicolumn{2}{c}{BadNets}  & \multicolumn{2}{c}{Blended} & \multicolumn{2}{c}{BadNets} & \multicolumn{2}{c}{Blended}                                                                                                                                                                         \\ \midrule
            $\sigma$ $\downarrow$ & BA                           & ASR                          & BA                           & ASR                         & BA                          & ASR                         & BA                        & ASR                       & BA                        & ASR                       & BA                        & ASR                       \\ \midrule
            0                                            & 89.78                        & 100                          & 90.03                        & 92.64                       & 56.04                       & 97.01                       & \textbf{55.12}            & \textbf{84.17}            & 85.96                     & 99.91                     & 85.34                     & 97.68                     \\ \midrule
            0.005                                        & 89.38                        & 100                          & 89.48                        & 91.15                       & 55.67                       & 95.44                       & 55.92                     & 85.99                     & 84.64                     & 99.91                     & 84.97                     & 97.29                     \\ \midrule
            0.01                                         & 88.85                        & 100                          & 88.67                        & 91.50                       & 57.19                       & 93.10                       & 55.46                     & 84.79                     & 83.12                     & 99.87                     & \textbf{81.90}            & \textbf{93.63}            \\ \midrule
            0.05                                         & 81.04                        & 100                          & 80.85                        & 81.03                       & 55.04                       & 87.26                       & 54.22                     & 86.06                     & \textbf{60.83}            & \textbf{6.61}             & 58.02                     & 98.48                     \\ \midrule
            0.1                                          & 69.77                        & 100                          & 68.10                        & 72.47                       & \textbf{51.43}              & \textbf{25.40}              & 50.12                     & 89.35                     & 43.55                     & 17.80                     & 43.26                     & 100                       \\ \midrule
            0.3                                          & 52.48                        & \multicolumn{1}{l}{99.99}    & \multicolumn{1}{l}{\textbf{53.05}}    & \multicolumn{1}{l}{\textbf{20.68}}   & \multicolumn{1}{l}{39.33}   & \multicolumn{1}{l}{7.49}    & \multicolumn{1}{l}{36.97} & \multicolumn{1}{l}{87.13} & \multicolumn{1}{l}{17.59} & \multicolumn{1}{l}{22.08} & \multicolumn{1}{l}{13.32} & \multicolumn{1}{l}{53.20} \\ \midrule
            0.5                                          & \textbf{44.24}               & \multicolumn{1}{l}{\textbf{14.56}}    & \multicolumn{1}{l}{45.26}    & \multicolumn{1}{l}{14.19}   & \multicolumn{1}{l}{32.26}   & \multicolumn{1}{l}{14.46}   & \multicolumn{1}{l}{31.60} & \multicolumn{1}{l}{84.23} & \multicolumn{1}{l}{10.17} & \multicolumn{1}{l}{48.90} & \multicolumn{1}{l}{9.27}  & \multicolumn{1}{l}{53.50} \\ \midrule
            0.7                                          & 40.37                        & \multicolumn{1}{l}{21.23}    & \multicolumn{1}{l}{39.71}    & \multicolumn{1}{l}{15.49}   & \multicolumn{1}{l}{29.06}   & \multicolumn{1}{l}{21.46}   & \multicolumn{1}{l}{25.38} & \multicolumn{1}{l}{88.69} & \multicolumn{1}{l}{8.74}  & \multicolumn{1}{l}{41.55} & \multicolumn{1}{l}{6.55}  & \multicolumn{1}{l}{61.50} \\ \midrule
            1                                            & 35.16                        & \multicolumn{1}{l}{32.27}    & \multicolumn{1}{l}{36.51}    & \multicolumn{1}{l}{29.24}   & \multicolumn{1}{l}{23.70}   & \multicolumn{1}{l}{31.30}   & \multicolumn{1}{l}{22.27} & \multicolumn{1}{l}{82.23} & \multicolumn{1}{l}{6.64}  & \multicolumn{1}{l}{31.59} & \multicolumn{1}{l}{6.46}  & \multicolumn{1}{l}{21.43} \\ \bottomrule
      \end{tabular}
      }
      \label{tab:DPSGD_search1}
      \vspace{-1.3em}
\end{table*}

\begin{table*}[!ht]
      \centering
      \caption{Results of DPSGD against the label-consistent attack with different noise scale $\sigma$. }
      \scalebox{0.85}{
      \begin{tabular}{@{}cccllccccccc@{}}
            \toprule
            $\epsilon$ $\downarrow$ & Poisoning Rate $\downarrow$ & \begin{tabular}[c]{@{}c@{}}$\sigma$ $\rightarrow$\\ Metric $\downarrow$\end{tabular} & \multicolumn{1}{c}{0}     & \multicolumn{1}{c}{0.005} & 0.01  & 0.05           & 0.1            & 0.3   & 0.5            & 0.7   & 1     \\ \midrule
            \multirow{4}{*}{16}     & \multirow{2}{*}{0.6\%}    & BA                         & 89.69                     & 89.62                     & 89.15 & \textbf{82.25} & 70.24          & 53.41 & 43.82          & 39.98 & 36.65 \\ \cmidrule(l){3-12}
                                    &                          & ASR                        & 88.79                     & 89.00                     & 88.16 & \textbf{4.30}  & 6.68           & 8.80  & 3.51           & 7.97  & 16.04 \\ \cmidrule(l){2-12}
                                    & \multirow{2}{*}{2.5\%}    & BA                         & 89.40                     & 89.56                     & 88.59 & 81.25          & 69.87          & 53.16 & \textbf{44.28} & 35.08 & 36.28 \\ \cmidrule(l){3-12}
                                    &                          & ASR                        & 97.95                     & 98.51                     & 98.57 & 99.42          & 80.32          & 69.04 & \textbf{7.97}  & 13.98 & 16.27 \\ \midrule
            \multirow{4}{*}{32}     & \multirow{2}{*}{0.6\%}    & BA                         & 89.75                     & 89.43                     & 88.77 & \textbf{82.27} & 71.12          & 53.42 & 47.52          & 39.18 & 37.06 \\ \cmidrule(l){3-12}
                                    &                          & ASR                        & 98.29                     & 99.02                     & 98.38 & \textbf{9.23}  & 5.76           & 13.60 & 6.29           & 15.64 & 14.70 \\ \cmidrule(l){2-12}
                                    & \multirow{2}{*}{2.5\%}    & BA                         & \multicolumn{1}{c}{89.87} & \multicolumn{1}{c}{89.38} & 89.09 & 82.15          & \textbf{69.79} & 53.02 & 44.08          & 38.31 & 36.13 \\ \cmidrule(l){3-12}
                                    &                          & ASR                        & \multicolumn{1}{c}{99.69} & \multicolumn{1}{c}{98.89} & 99.57 & 99.86          & \textbf{9.06}  & 18.88 & 6.62           & 20.97 & 7.32  \\ \bottomrule
      \end{tabular}
      }
      \label{tab:DPSGD_search2}
      \vspace{-1.3em}
\end{table*}

\begin{table*}[!ht]
   \centering
   \caption{Results of DPSGD against the WaNet with different noise scale $\sigma$. }
   \scalebox{0.85}{
     \begin{tabular}{ccccccc}
         \toprule
         Dataset $\rightarrow$                        & \multicolumn{2}{c}{CIFAR-10}  & \multicolumn{2}{c}{ImageNet} & \multicolumn{2}{c}{VGGFace2}                                                                     \\ \midrule
         $\sigma$ $\downarrow$, Metric $\rightarrow$                                           & BA                              & ASR                             & BA             & ASR            & BA             & ASR            \\ \midrule
         0                                            & 87.76                                        & 87.06                                         & 53.26          & 96.06          & 83.62          & 95.22          \\ \midrule
         0.005                                        & 88.21                                        & 84.71                                         & 53.38          & 98.14          & 80.72          & 96.47          \\ \midrule
         0.01                                         & 88.13                                        & 82.05                                         & 54.03          & 97.72          & 79.88          & 83.57          \\ \midrule
         0.05                                         & 81.03                                        & 80.59                                         & 51.48          & 97.00          & \textbf{54.04} & \textbf{30.85} \\ \midrule
         0.1                                          & 66.85                                        & 72.75                                         & 47.65          & 98.63          & 33.89          & 99.82          \\ \midrule
         0.3                                          & \textbf{52.13}                               & \textbf{39.78}                                & 29.65          & 96.45          & 9.44           & 73.99          \\ \midrule
         0.5                                          & 41.56                                        & 39.11                                         & 22.13          & 93.45          & 6.41           & 80.82          \\ \midrule
         0.7                                          & 36.51                                        & 33.60                                         & \textbf{20.72} & \textbf{54.12} & 6.10           & 86.98          \\ \midrule
         1                                            & 32.66                                        & 39.75                                         & 15.15          & 58.10          & 6.36           & 52.16          \\ \midrule
      \end{tabular}
   }
   \label{tab:DPSGD_search3}
   \vspace{-1em}
\end{table*}

\noindent \textbf{Results.}
As shown in Table \ref{tab:main_vgg}, our defense still reaches the best performance even compared with NC and NAD. Specifically, the BA of NC is on par with that of our method whereas it is with the sacrifice of ASR. These results verify the effectiveness of our defense again.

\section{Searching Best Results for DPSGD and NAD}
\label{sec:search_best}
The effectiveness of DPSGD and NAD is sensitive to their hyper-parameters. Here we search for their best results based on the criteria that `BA $-$ ASR' reaches the highest value after the defense.

\subsection{Searching Best Results for DPSGD}
In general, the larger the $\sigma$, the smaller the ASR while also the smaller the BA. 
The results of DPSGD are shown in Table \ref{tab:DPSGD_search1}-\ref{tab:DPSGD_search3}, where the best results are marked in boldface.

\begin{table*}[!ht]
      \centering
      \vspace{-1em}
      \caption{Results of the fine-tuning process in NAD against the BadNets and blended attack with different learning rates $\eta$. }
      \scalebox{0.83}{
      \begin{tabular}{@{}ccccccccccccc@{}}
            \toprule
            Dataset $\rightarrow$                     & \multicolumn{4}{c}{CIFAR-10} & \multicolumn{4}{c}{ImageNet} & \multicolumn{4}{c}{VGGFace2}                                                                                                                                           \\ \midrule
            Attack $\rightarrow$                      & \multicolumn{2}{c}{BadNets}  & \multicolumn{2}{c}{Blended}  & \multicolumn{2}{c}{BadNets}  & \multicolumn{2}{c}{Blended} & \multicolumn{2}{c}{BadNets} & \multicolumn{2}{c}{Blended}                                                 \\ \midrule
            $\eta$ $\downarrow$, Metric $\rightarrow$ & BA                           & ASR                          & BA                           & ASR                         & BA                          & ASR                         & BA    & ASR   & BA    & ASR   & BA    & ASR   \\ \midrule
            0.1                                       & 32.64                        & 5.12                         & 38.24                        & 1.43                        & \textbf{62.73}                       & \textbf{1.09}                        & \textbf{56.72} & \textbf{0.95}  & \textbf{90.35} & \textbf{24.36} & 91.39 & 99.96 \\ \midrule
            0.01                                      & \textbf{90.70}                        & \textbf{15.66}                        & \textbf{89.77}                        & \textbf{11.32}                       & 80.37                       & 43.90                       & 80.28 & 95.82 & 92.04 & 83.80 & \textbf{92.95} & \textbf{99.96} \\ \midrule
            0.001                                     & 94.91                        & 100                          & 93.25                        & 96.79                       & 81.47                       & 86.79                       & 80.85 & 99.12 & 91.49 & 90.91 & 92.91 & 99.96 \\ \bottomrule
      \end{tabular}
      }
      \label{tab:nad_lr1}
\end{table*}

\begin{table*}[!ht]
      \centering
      \caption{Results of the fine-tuning process in NAD against the label-consistent attack with different learning rates $\eta$ on the CIFAR-10 dataset. }
      \scalebox{0.9}{
      \begin{tabular}{@{}cccccc@{}}
            \toprule
            $\epsilon$ $\downarrow$ & Poisoning Rate $\downarrow$ & \begin{tabular}[c]{@{}c@{}}$\eta$ $\rightarrow$\\ Metric $\downarrow$\end{tabular} & 0.1   & 0.01  & 0.001 \\ \midrule
            \multirow{4}{*}{16}     & \multirow{2}{*}{0.6\%}    & BA                         & 36.77 & \textbf{89.91} & 95.38 \\ \cmidrule(l){3-6}
                                    &                          & ASR                        & 0.63  & \textbf{44.90} & 94.69 \\ \cmidrule(l){2-6}
                                    & \multirow{2}{*}{2.5\%}    & BA                         & \textbf{31.12} & 93.07 & 95.03 \\ \cmidrule(l){3-6}
                                    &                          & ASR                        & \textbf{0}     & 99.67 & 100   \\ \midrule
            \multirow{4}{*}{32}     & \multirow{2}{*}{0.6\%}    & BA                         & 30.73 & \textbf{91.91} & 94.89 \\ \cmidrule(l){3-6}
                                    &                          & ASR                        & 1.15  & \textbf{38.27} & 99.07 \\ \cmidrule(l){2-6}
                                    & \multirow{2}{*}{2.5\%}    & BA                         & \textbf{35.93} & 91.23 & 94.79 \\ \cmidrule(l){3-6}
                                    &                          & ASR                        & \textbf{0}     & 96.19 & 99.98 \\ \bottomrule
      \end{tabular}
      }
      \label{tab:nad_lr2}
\end{table*}

\begin{table*}[!ht]
   \centering
   \caption{Results of the fine-tuning process in NAD against WaNet with different learning rates $\eta$. }
   \scalebox{0.9}{
      \begin{tabular}{c|cc|cc|cc}
         \toprule
         Dataset $\rightarrow$                     & \multicolumn{2}{c}{CIFAR-10} & \multicolumn{2}{c}{ImageNet} & \multicolumn{2}{c}{VGGFace2}                         \\ \hline
         $\eta$ $\downarrow$, Metric $\rightarrow$ & BA                           & ASR                          & BA                           & ASR   & BA    & ASR   \\ \hline
         0.1                                       & 48.74                        & 2.75                         & \textbf{59.11} & \textbf{0.34}  & \textbf{89.05} & \textbf{91.40} \\ \hline
         0.01                                      & \textbf{90.60} & \textbf{10.37} & 79.11                        & 67.09 & 89.01 & 99.69 \\ \hline
         0.001                                     & 94.16                        & 97.65                        & 80.63                        & 99.39 & 88.19 & 99.73 \\ \bottomrule
      \end{tabular}
   }
   \label{tab:nad_lr3}
   \vspace{-1em}
\end{table*}

\begin{table*}[!ht]
   \centering
   \vspace{-1em}
   \caption{Results of the NAD against attacks on the CIFAR-10 dataset with different $\beta$. }
   \scalebox{0.95}{
      \begin{tabular}{@{}cccccccc@{}}
         \toprule
         Attack $\downarrow$             & \begin{tabular}[c]{@{}c@{}}$\beta$ $\rightarrow$\\ Metric $\downarrow$\end{tabular} & 500   & 1,000          & 1,500          & 2,000 & 2,500 & 5,000 \\ \midrule
         \multirow{2}{*}{BadNets}        & BA                                                                                  & 92.96 & 91.87          & \textbf{90.47} & 86.26 & 65.26 & 19.91 \\ \cmidrule(l){2-8}
                                         & ASR                                                                                 & 27.74 & 7.92           & \textbf{1.20}  & 1.74  & 5.81  & 0.40  \\ \midrule
         \multirow{2}{*}{Blended Attack} & BA                                                                                  & 91.14 & \textbf{89.72} & 85.30          & 75.43 & 61.18 & 26.87 \\ \cmidrule(l){2-8}
                                         & ASR                                                                                 & 6.25  & \textbf{1.51}  & 3.19           & 3.55  & 3.67  & 9.61  \\ \midrule
         \multirow{2}{*}{WaNet}          & BA                                                                                  & 92.43 & \textbf{91.83} & 88.08          & 78.22 & 58.32 & 26.63 \\ \cmidrule(l){2-8}
                                         & ASR                                                                                 & 18.48 & \textbf{9.49}  & 3.58           & 3.54  & 3.12  & 0.32  \\ \bottomrule
      \end{tabular}
   }
   \label{tab:NAD_beta1}
\end{table*}

\subsection{Searching Best Results for NAD}

We found that the fine-tuning stage of NAD is sensitive to the learning rate. We search the best initial learning rate from $\{0.1, 0.01, 0.001\}$. As shown in Table \ref{tab:nad_lr1}-\ref{tab:nad_lr3}, a very large learning rate significantly reduces the BA, while a very small learning rate can not reduce the ASR effectively. To keep a relatively large BA while maintaining a small ASR, we set $\eta=0.01$ in the fine-tuning stage.

The distillation stage of NAD is also sensitive to its hyper-parameter $\beta$. We select the best $\beta$ via the grid-search. The results are shown in Table \ref{tab:NAD_beta1}-\ref{tab:NAD_beta4}.

\begin{table*}[!t]
   \centering
   \caption{Results of the NAD against attacks on the ImageNet dataset with different $\beta$. }
   \scalebox{0.95}{
      \begin{tabular}{@{}cccccccc@{}}
         \toprule
         Attack $\downarrow$             & \begin{tabular}[c]{@{}c@{}}$\beta$ $\rightarrow$\\ Metric $\downarrow$\end{tabular} & 5,000 & 6,000 & 7,000 & 8,000          & 9,000          & 10,000 \\ \midrule
         \multirow{2}{*}{BadNets}        & BA                                                                                  & 79.98 & 79.20 & 76.88 & \textbf{73.49} & 69.46          & 67.55  \\ \cmidrule(l){2-8}
                                         & ASR                                                                                 & 24.65 & 15.50 & 16.99 & \textbf{5.81}  & 5.95           & 2.51   \\ \midrule
         \multirow{2}{*}{Blended} & BA                                                                                  & 79.98 & 77.57 & 74.88 & \textbf{70.88} & 64.44          & 53.75  \\ \cmidrule(l){2-8}
                                         & ASR                                                                                 & 88.34 & 75.79 & 15.12 & \textbf{8.15}  & 3.47           & 2.99   \\ \midrule
         \multirow{2}{*}{WaNet}          & BA                                                                                  & 79.65 & 79.06 & 77.63 & 76.77          & \textbf{75.27} & 74.88  \\ \cmidrule(l){2-8}
                                         & ASR                                                                                 & 76.06 & 69.23 & 61.54 & 45.97          & \textbf{31.43} & 28.58  \\ \bottomrule
      \end{tabular}
   }
   \label{tab:NAD_beta2}
\end{table*}

\begin{table*}[!ht]
   \centering
   \caption{Results of the NAD against attacks on the VGGFace2 dataset with different $\beta$. }
   \scalebox{0.95}{
      \begin{tabular}{cccccccc}
         \toprule
         Attack $\downarrow$             & \begin{tabular}[c]{@{}c@{}}$\beta$ $\rightarrow$\\ Metric $\downarrow$\end{tabular} & 10    & 30             & 40             & 50    & 60             & 70  \\ \midrule
         \multirow{2}{*}{BadNets}        & BA                                                                                  & 89.79 & 78.10          & 76.62          & 83.82 & \textbf{87.29} & 47.53       \\ \cmidrule(l){2-8}
                                         & ASR                                                                                 & 36.02 & 17.87          & 0.18           & 14.52 & \textbf{9.49}  & 0              \\ \midrule
         \multirow{2}{*}{Blended} & BA                                                                                  & 90.18 & 83.48          & \textbf{85.27} & 73.43 & 75.01          & 51.81    \\ \cmidrule(l){2-8}
                                         & ASR                                                                                 & 97.11 & 68.97          & \textbf{10.25} & 0     & 0.36           & 0.27        \\ \midrule
         \multirow{2}{*}{WaNet}          & BA                                                                                  & 82.98 & \textbf{80.69} & 84.27          & 26.87 & 20.71          & 75.10\\ \cmidrule(l){2-8}
                                         & ASR                                                                                 & 66.10 & \textbf{4.41}  & 15.98          & 0.06  & 1.88           & 0.54 \\ \bottomrule
      \end{tabular}
   }
   \label{tab:NAD_beta3}
\end{table*}

\begin{table*}[!ht]
      \centering
      \caption{Results of the NAD against the label consistent attack on the CIFAR-10 dataset with different $\beta$. }
      \scalebox{0.95}{
      \begin{tabular}{@{}ccccccccc@{}}
            \toprule
            $\epsilon \downarrow$             & Poisoning Rate $\downarrow$           & \begin{tabular}[c]{@{}c@{}}$\beta$ $\rightarrow$\\ Metric $\downarrow$\end{tabular} & 500   & 1,000           & 1,500           & 2,000           & 2,500  & 5,000  \\ \midrule
            \multirow{4}{*}{16} & \multirow{2}{*}{0.6\%} & BA     & 92.94 & 91.66          & 90.29          & \textbf{85.14} & 73.05 & 18.49 \\ \cmidrule(l){3-9}
                                &                       & ASR    & 53.11 & 33.78          & 12.43          & \textbf{4.39}  & 4.36  & 0.1   \\ \cmidrule(l){2-9}
                                & \multirow{2}{*}{2.5\%} & BA     & 93.36 & \textbf{93.04} & 90.31          & 75.11          & 54.55 & 24.37 \\ \cmidrule(l){3-9}
                                &                       & ASR    & 36.13 & \textbf{3.95}  & 1.88           & 3.13           & 0.22  & 0     \\ \midrule
            \multirow{4}{*}{32} & \multirow{2}{*}{0.6\%} & BA     & 92.81 & 92.13          & 91.21          & \textbf{85.72} & 60.06 & 18.81 \\ \cmidrule(l){3-9}
                                &                       & ASR    & 52.50 & 57.63          & 38.53          & \textbf{9.99}  & 1.33  & 0.06  \\ \cmidrule(l){2-9}
                                & \multirow{2}{*}{2.5\%} & BA     & 93.40 & 92.02          & \textbf{90.98} & 86.32          & 68.45 & 20.53 \\ \cmidrule(l){3-9}
                                &                       & ASR    & 48.73 & 17.99          & \textbf{6.54}  & 2.93           & 1.79  & 0     \\ \bottomrule
      \end{tabular}
      }
      \label{tab:NAD_beta4}
      \vspace{-0.8em}
\end{table*}

\section{Defending against Label-Consistent Attack with a Smaller Poisoning Rate}
\label{sec:smaller_p}

For the label-consistent attack, except for the 2.5\% poisoning rate examined in the main manuscript, 0.6\% is also an important setting provided in its original paper \citep{turner2019label}. In this section, we compare different defenses against the label-consistent attack with poisoning rate $\gamma = 0.6\%$.

As shown in Table \ref{tab:consistent_new}, when defending against label-consistent attack with a 0.6\% poisoning rate, our method is still significantly better than defenses having the same requirements ($i.e.$, DPSGD and ShrinkPad). Even compared with those having the additional requirement ($i.e.$, NC and NAD) under their best settings, our defense is still better or on par with them under the default settings. These results verify the effectiveness of our method again.

\begin{table*}[ht]
\vspace{-1em}

      \centering
      \caption{The effectiveness (\%) of defending against the label-consistent
            attack with 0.6\% and 2.5\% poisoning rate on CIFAR-10 dataset. Among all defense methods, the one with the best performance is indicated in boldface and the value with underline denotes the second-best result. Note that NC and NAD require to have an additional local benign dataset, which is not required in DPSGD, ShrinkPad, and our method.}
    \scalebox{0.91}{
      \begin{tabular}{ccc|c|cc|ccc}
            \toprule
            $\epsilon$ $\downarrow$ & poison rate $\downarrow$ & \begin{tabular}[c]{@{}c@{}}Defense $\rightarrow$\\ Metric $\downarrow$\end{tabular} & No Defense & NC    & NAD   & DPSGD & ShrinkPad & DBD (ours) \\ \hline
            \multirow{4}{*}{16}     & \multirow{2}{*}{0.6\%}    & BA                         & 95.34  & \textbf{94.51} & 85.14 & 82.25 & 90.59     & \underline{91.88}      \\ \cline{3-9}
                                    &                          & ASR                        & 92.20  & \underline{0.73}  & 4.39  & 4.30  & 10.22     & \textbf{0.46}       \\ \cline{2-9}
                                    & \multirow{2}{*}{2.5\%}    & BA                         & 94.90  & \textbf{94.94} & \underline{93.04} & 44.28 & 90.67     & 89.67      \\ \cline{3-9}
                                    &                          & ASR                        & 99.33  & \underline{1.31}  & 3.95  & 7.97  & 9.60      & \textbf{0.01}       \\ \hline
            \multirow{4}{*}{32}     & \multirow{2}{*}{0.6\%}    & BA                         & 94.94  & \textbf{94.53} & 85.72 & 88.27 & \underline{90.48}     & 90.04      \\ \cline{3-9}
                                    &                          & ASR                        & 97.69  & \underline{0.72}  & 9.99  & 9.23  & 11.47     & \textbf{0.15}       \\ \cline{2-9}
                                    & \multirow{2}{*}{2.5\%}    & BA                         & 94.82  & \textbf{94.57} & 90.98 & 69.79 & 90.83     & \underline{91.45}      \\ \cline{3-9}
                                    &                          & ASR                        & 99.19  & \underline{1.36}  & 6.54  & 9.06  & 13.03     & \textbf{0.34}       \\ \bottomrule
      \end{tabular}
      }
      \label{tab:consistent_new}
\end{table*}

\begin{table}[ht]
   \centering
   \caption{DBD against BadNets with different triggers on CIFAR-10 dataset.}
   \scalebox{0.89}{
      \begin{tabular}{c|c|ccccc|cccc}
         \toprule
         \multicolumn{2}{c|}{\multirow{2}{*}{}} & \multicolumn{5}{c|}{Location} & \multicolumn{4}{c}{Trigger Size}                                                                 \\ \cline{3-11}
         \multicolumn{2}{c|}{}                  & \tabincell{c}{Upper                                                                                                              \\Left} & \tabincell{c}{Upper\\Right} & Center & \tabincell{c}{Lower\\Left} & \tabincell{c}{Lower\\Right} & $1 \times 1$ & $2 \times 2$ & $3 \times 3$ & $4 \times 4$ \\ \hline
         \multirow{2}{*}{\tabincell{c}{No                                                                                                                                          \\Defense}}    & BA    &  95.01          &   95.01          &  94.97      &  94.72          & 94.77            & 94.70             & 94.99             & 95.01             &  94.89            \\
                                                & ASR                           & 99.78                            & 99.70 & 99.86 & 99.68 & 99.70 & 92.13 & 99.35 & 99.78 & 99.86 \\ \hline
         \multirow{2}{*}{\tabincell{c}{DBD                                                                                                                                         \\(ours)}} & BA & 92.46 & 92.47 & 93.32 & 92.92 & 92.58 & 93.10 & 92.91 & 92.46 & 92.46 \\
                                                & ASR                           & 1.27                             & 0.53  & 0.51  & 0.64  & 1.41  & 0.80  & 0.65  & 1.27  & 1.02  \\ \bottomrule
      \end{tabular}
   }
   \label{tab:diff_triggers}
\end{table}

\section{Defending against Attacks with Different Trigger Patterns}
\label{sec:diff_trigger}

In this section, we verify whether DBD is still effective when different trigger patterns are adopted.

\noindent \textbf{Settings.}
For simplicity, we adopt the BadNets on the CIFAR-10 dataset as an example for the discussion. Specifically, we change the \emph{location} and \emph{size} of the backdoor trigger while keeping other settings unchanged to evaluate the BA and ASR before and after our defense.

\noindent \textbf{Results.}
As shown in Table \ref{tab:diff_triggers}, although there are some fluctuations, the ASR is smaller than $2\%$ while the BA is greater than $92\%$ in every cases. In other words, our method is effective in defending against attacks with different trigger patterns.

\section{Defending against Attacks with Dynamic Triggers}
In this section, we verify whether DBD is still effective when attackers adopt dynamic triggers.

\noindent \textbf{Settings.}
We compare DBD with MESA \citep{qiao2019defending} in defending the dynamic attack discussed in \citep{qiao2019defending} on the CIFAR-10 dataset as an example for the discussion. This dynamic attack uses a distribution of triggers instead of a fixed trigger.

\noindent \textbf{Results.}
The BA and ASR of DBD are 92.4\% and 0.4\%, while those
of MESA are 94.8\% and 2.4\%. However, we find MESA
failed in defending against blended attack (for it can not
correctly detect the trigger) whereas DBD is still effective.
These results verified the effectiveness of our defense.

\comment{

\section{Defending against Attacks with Different Target Labels}
\label{sec:diff_label}

In this section, we verify whether DBD is still effective when different target labels are adopted.

\noindent \textbf{Settings.}
For simplicity, we conduct evaluations on CIFAR-10 dataset as an example for the discussion. Except for the target label, other settings are the same as those used in Section \ref{sec:main}.

\noindent \textbf{Results.}
As shown in Table \ref{tab:diff_target}, although there are some fluctuations, our method is still effective in defending against all attacks in each case, $i.e.$, our method is effective in defending against attacks with different target labels.

}

\section{Discussions}
\subsection{Effects of Hyper-parameters}

\noindent \textbf{Settings.}
Here we analyze the effect of filtering rate $\alpha$, which is the only key method-related hyper-parameter in our DBD. We adopt the results on the CIFAR-10 dataset for discussion. Except for the studied parameter $\alpha$, other settings are the same as those used in Section \ref{sec:main}.

\begin{figure}[ht]
\begin{minipage}[b]{0.48\linewidth}
    \centering
    \includegraphics[width=\textwidth]{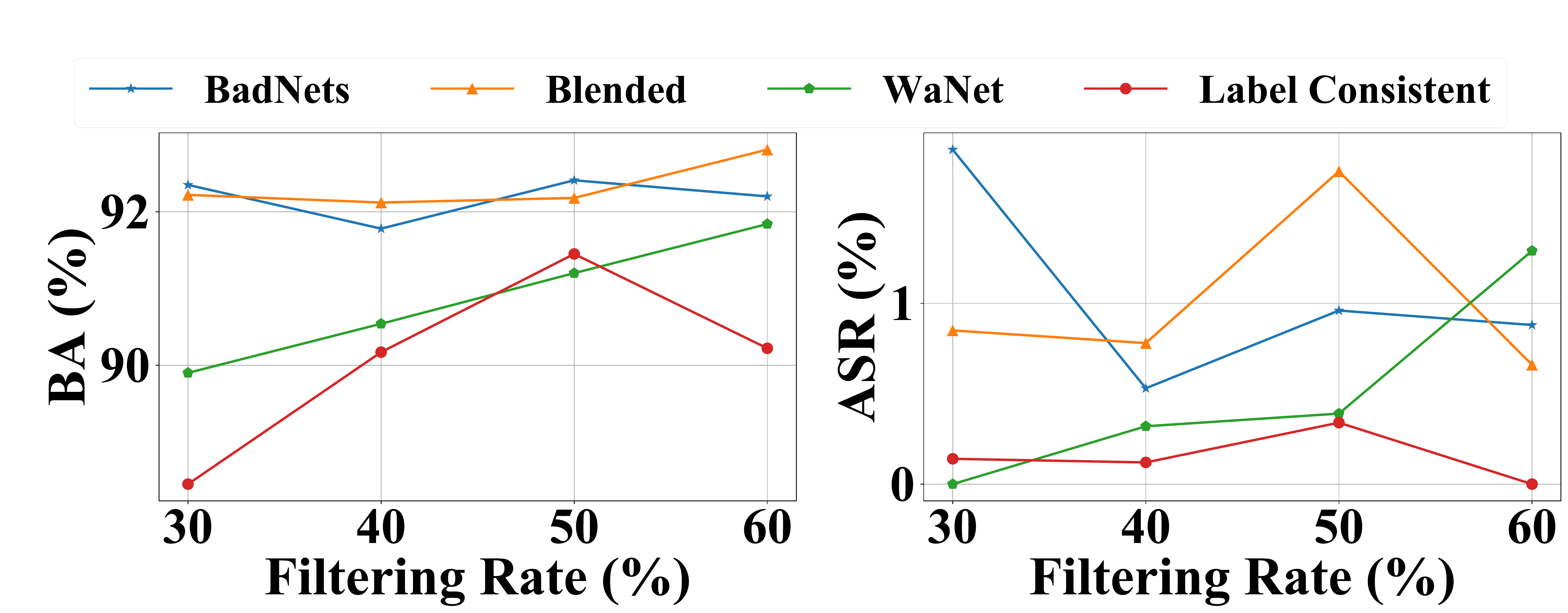}
    \vspace{-1.2em}
    \caption{The effects of filtering rate. }
    \label{fig:alpha}
\end{minipage}\quad
\begin{minipage}[b]{0.48\linewidth}
    \centering
    \includegraphics[width=\textwidth]{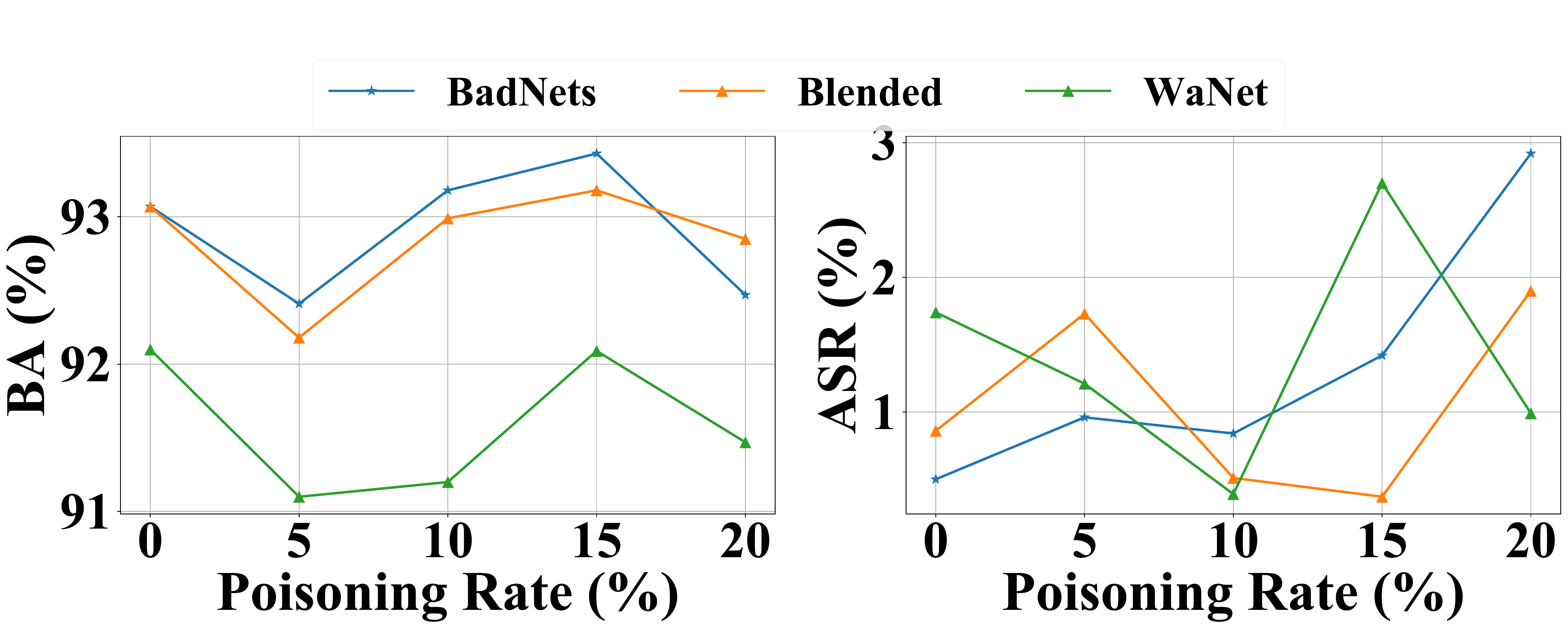}
    \vspace{-1.2em}
    \caption{The effects of poisoning rate. }
    \label{fig:gamma}
\end{minipage}
\end{figure}

\vspace{0.3em}
\noindent \textbf{Results.}
The number of labeled samples used in the third stage increase with the increase of filtering rate $\alpha$, while the probability that the filtered high-credible dataset contains poisoned samples also increases. As shown in Figure \ref{fig:alpha}, DBD can still maintain relatively high benign accuracy even when the filtering rate $\alpha$ is relatively small ($e.g.$, 30\%). It is mostly due to the high-quality of learned purified feature extractor and the semi-supervised fine-tuning process. 
DBD can also reach a nearly 0\% attack success rate in all cases. However, we also have to notice that the high-credible dataset may contain poisoned samples when $\alpha$ is very large, which in turn creates hidden backdoors again during the fine-tuning process. Defenders should specify $\alpha$ based on their specific needs.

\subsection{Defending Attacks with Various Poisoning Rates}
\noindent \textbf{Settings. } 
We evaluate our method in defending against attacks with different poisoning rate $\gamma$ on CIFAR-10 dataset. Except for $\gamma$, other settings are the same as those used in Section \ref{sec:main}.

\section{More Details about SimCLR, SCE, and MixMatch}
\label{sec:more_loss}

\noindent \textbf{NT-Xent Loss in SimCLR.}
Given a sample mini-batch containing $N$ different samples, SimCLR first applies two separate data augmentations toward each sample to obtain $2N$ augmented samples. The loss for a positive pair of sample $(i, j)$ can be defined as:
 \begin{equation}
 \mathcal{L}_{i, j}=-\log \frac{\exp \left(\operatorname{sim}\left(\boldsymbol{z}_{i}, \boldsymbol{z}_{j}\right) / \tau\right)}{\sum_{k=1}^{2 N} \mathbb{I}{\{k \neq i\}} \cdot \exp \left(\operatorname{sim}\left(\boldsymbol{z}_{i}, \boldsymbol{z}_{k}\right) / \tau\right)},
 \end{equation}
where $\operatorname{sim}(\cdot, \cdot)$ is the cosine similarity, $z_i$ is the feature representation of sample $i$, $\tau$ is the temperature parameter, and $\mathbb{I}{\{k \neq i\}} \in \{0,1\}$ indicating whether $k \neq i$. The NT-Xent Loss is computed across all $2N$ positive pairs in this mini-batch.

\noindent \textbf{SCE.}
The symmetric cross entropy (SCE) can be defined as:
\begin{equation}
    \mathcal{L}_{SCE}=H(p, q)+H(q, p),
\end{equation}
where $H(p, q)$ is the cross entropy, $H(q, p)$ is the reversed cross entropy, $p$ is the prediction, and $q$ is the one-hot label (of the evaluated sample).

\noindent \textbf{MixMatch Loss.}
For a batch $\mathcal{X}$ of labeled samples and a batch $\mathcal{U}$ of unlabeled samples ($|\mathcal{X}|=|\mathcal{U}|$), MixMatch produces a guessed label $\bar{q}$ for each unlabled sample $u\in \mathcal{U}$ and applies MixUp \citep{zhang2018mixup} to obtain the augmented $\mathcal{X}^{'}$ and  $\mathcal{U}^{'}$. The loss $\mathcal{L}_{\mathcal{X}}$ and $\mathcal{L}_{\mathcal{U}}$ can be defined as:
\begin{equation}
    \mathcal{L}_{\mathcal{X}}=\frac{1}{\left|\mathcal{X}^{\prime}\right|} \sum_{(x, q) \in \mathcal{X}^{\prime}} H\left(p_{x}, q\right),
\end{equation}
where $p_{x}$ is the prediction of $x$, $q$ is its one-hot label, and $H(\cdot, \cdot)$ is the cross entropy.
\begin{equation}
    \mathcal{L}_{\mathcal{U}}=\frac{1}{K \cdot \left|\mathcal{U}^{\prime}\right|} \sum_{(u, \bar{q}) \in \mathcal{U}^{\prime}}\left\|p_{u}-\bar{q}\right\|_{2}^{2},
\end{equation}
where $p_{u}$ is the prediction of $u$, $\bar{q}$ is its guessed one-hot label, and $K$ is the number of classes.

By combining $\mathcal{L}_{\mathcal{X}}$ with $\mathcal{L}_{\mathcal{U}}$, the MixMatch loss can be defined as:
\begin{equation}
    \mathcal{L}=\mathcal{L}_{\mathcal{X}}+\lambda_{\mathcal{U}} \cdot \mathcal{L}_{\mathcal{U}},
\end{equation}
where $\lambda_{\mathcal{U}}$ is a hyper-parameter.

\section{Computational Facilities}
\label{sec:train_f}

We conduct all experiments on two Ubuntu 18.04 servers having different GPUs. One has four NVIDIA GeForce RTX 2080 Ti GPUs with 11GB memory (dubbed `RTX 2080Ti') and the another has three NVIDIA Tesla V100 GPUs with 32GB memory (dubbed `V100').

\noindent \textbf{Computational Facilities for Attacks.}
All experiments are conducted with a single RTX 2080 Ti.

\noindent \textbf{Computational Facilities for Defenses.}
Since we do not use a memory-efficient implementation of DenseNet-121, we conduct DPSGD experiments on the VGGFace2 dataset with a single V100. Other experiments of baseline defenses are conducted with a single RTX 2080 Ti. For our defense, we adopt PyTorch \citep{paszke2019pytorch} distributed data-parallel and automatic mixed precision training \citep{micikevicius2018mixed} with two RTX 2080 Ti for self-supervised learning on the VGGFace2 dataset. Other experiments are conducted with a single RTX 2080 Ti.

\section{Computational Cost}
\label{sec: computational cost}
In this section, we analyze the computational cost of our method stage by stage, compared to standrad supervised learning.

\noindent \textbf{Stage 1.}
Self-supervised learning is known to have a higher computational cost than standard supervised learning \citep{chen2020simple,he2020momentum}. In our experiments, SimCLR requires roughly four times the computational cost of standard supervised learning. Since we intend to get a purified instead of well-trained feature extractor,  we train the feature extractor ($i.e.$, backbone) lesser epochs than the original SimCLR to reduce the training time. As described in Section \ref{sec:DBD_set}, we find 100 epochs is enough to preserve effectiveness.

\noindent \textbf{Stage 2.}
Since we freeze the backbone and only train the remaining fully connected layers, the computational cost is roughly 60\% of standard supervised learning.

\noindent \textbf{Stage 3.}
Semi-supervised learning is known to have a extra labeling cost compared with standard supervised learning \citep{gao2020consistency}. In our experiments, MixMatch requires roughly two times the computation cost of standard supervised learning.

We will explore a more computational efficient training method in our future work.

\section{Comparing our DBD with Detection-based Backdoor Defenses}
In this paper, we do not intend to filter malicious and benign samples accurately, as we mentioned in Section \ref{sec:filtering}. However, we notice that the second stage of our DBD can serve as a detection-based backdoor defense for it can filter poisoned samples. In this section, we compare the filtering ability of our DBD (stage 2) with existing detection-based backdoor defenses.

\noindent \textbf{Settings.}
We compare our DBD with two representative detection-based methods, including, Spectral Signatures (SS) \citep{tran2018spectral} and Activation Clustering (AC) \citep{chen2019detecting}, on the CIFAR-10 dataset. These detection-based methods ($e.g.$, SS and AC) filter malicious samples from the training set and train the model on non-malicious samples. Specifically, we re-implement SS in PyTorch based on its official code\footnote{\url{https://github.com/MadryLab/backdoor_data_poisoning}} and adopt the open-source code\footnote{\url{https://github.com/ain-soph/trojanzoo/blob/main/trojanvision/defenses/backdoor/activation_clustering.py}} for AC, following the settings in their original paper. In particular, since SS filters $1.5 \varepsilon$ malicious samples for each class, where $\varepsilon$ is the key hyper-parameter means the upper bound of the number of poisoned training samples, we adopt different $\varepsilon$ for a fair comparison.

\noindent \textbf{Results.}
As shown in Table \ref{tab:SS_DBD}-\ref{tab:AC_DBD}, the filtering performance of DBD is on par with that of SS and AC. DBD is even better than those methods when filtering poisoned samples generated by more complicated attacks ($i.e.$, WaNet and Label-Consistent). Besides, we also conduct the standard training on non-malicious samples filtered by SS and AC. As shown in Table \ref{tab:detection_performance}, the hidden backdoor will still be created in many cases, even though the detection-based defenses are sometimes accurate. This is mainly because these methods may not able to remove enough poisoned samples while preserving enough benign samples simultaneously, $i.e.$, there is a trade-off between BA and ASR.

\begin{table}[tb]
\vspace{-1.5em}
\centering
\caption{The successful filtering rate (\# filtered poisoned samples / \# all filtered samples, \%) under different $\varepsilon$ in the target class on CIFAR-10 dataset.}
\begin{tabular}{c|c|cccc}
\toprule
Attack $\downarrow$                            & Defense $\downarrow$, $\varepsilon \rightarrow$  & 250            & 500            & 1000           & 1500           \\ \hline
\multirow{2}{*}{BadNets}          & SS              & 95.73          & 93.20          & 87.60          & \textbf{80.71} \\
                                  & DBD             & \textbf{100}   & \textbf{97.60} & \textbf{90.87} & 70.09          \\ \hline
\multirow{2}{*}{Blended}          & SS              & 0.80           & 10.53          & 28.87          & 35.29          \\
                                  & DBD             & \textbf{97.87} & \textbf{94.67} & \textbf{87.27} & \textbf{75.16} \\ \hline
\multirow{2}{*}{WaNet}            & SS              & 0              & 0              & 1.00           & 7.42           \\
                                  & DBD             & \textbf{100}   & \textbf{100}   & \textbf{99.47} & \textbf{97.46} \\ \hline
\multirow{2}{*}{Label-Consistent} & SS              & 2.40           & 4.40           & 9.00           & 13.78          \\
                                  & DBD             & \textbf{43.47} & \textbf{37.07} & \textbf{34.47} & \textbf{32.53} \\ \bottomrule
\end{tabular}
\label{tab:SS_DBD}
\vspace{-0.5em}
\end{table}

\begin{table}[tb]
\centering
\caption{The number of remaining poisoned samples over filtered non-malicious samples on CIFAR-10 dataset. }
\begin{tabular}{c|cccc}
\toprule
Defense $\downarrow$, Attack $\rightarrow$ & BadNets    & Blended    & WaNet      & Label-Consistent \\ \hline
SS ($\varepsilon=500$)       & 1801/42500 & 2421/42500 & 2500/42500 & 1217/42500       \\
SS ($\varepsilon=1000$)       & 1186/35000 & 2067/35000 & 2400/35000 & 1115/35000       \\ \hline
AC        & 0/42500    & 0/37786    & 5000/45546 & 1250/39998       \\
DBD       & 8/25000    & 6/25000    & 38/25000   & 13/25000         \\ \bottomrule
\end{tabular}
\label{tab:AC_DBD}
\vspace{-0.5em}
\end{table}

\begin{table}[!t]
\centering
\caption{The BA (\%) and ASR (\%) of models trained on non-malicious samples filtered by SS and AC on CIFAR-10 dataset.}
\begin{tabular}{c|cc|cc|cc}
\toprule
Defense $\rightarrow$          & \multicolumn{2}{c|}{SS ($\varepsilon=500$)} & \multicolumn{2}{c|}{SS ($\varepsilon=1000$)} & \multicolumn{2}{c}{AC} \\ \hline
Attack $\downarrow$, Metric $\rightarrow$  & BA         & ASR        & BA         & ASR        & BA         & ASR       \\ \hline
BadNets          & 92.99      & 100        & 93.27      & 99.99      & 85.90      & 0         \\
Blended          & 92.84      & 99.07      & 92.56      & 99.18      & 77.17      & 0         \\
WaNet            & 92.69      & 98.13      & 91.92      & 99.00      & 84.60      & 99.02     \\
Label-Consistent & 92.93      & 99.79      & 92.88      & 99.86      & 75.95      & 99.75     \\ \bottomrule
\end{tabular}
\label{tab:detection_performance}
\end{table}

\begin{table}[!t]
\centering
\caption{The BA (\%) and ASR (\%) of our DBD defending against four attacks with different self-supervised methods on CIFAR-10 dataset.}
\begin{tabular}{c|cccccccc}
\toprule
Attack $\rightarrow$        & \multicolumn{2}{c}{BadNets} & \multicolumn{2}{c}{Blended} & \multicolumn{2}{c}{WaNet} & \multicolumn{2}{c}{Label-Consistent} \\ \hline
Method $\downarrow$, Metric $\rightarrow$ & BA            & ASR         & BA            & ASR         & BA           & ASR        & BA                & ASR              \\ \hline
SimCLR         & 92.41         & 0.96        & 92.18         & 1.73        & 91.20        & 0.39       & 91.45             & 0.34             \\ \hline
MoCo-V2        & 93.01         & 1.21        & 92.42         & 0.24        & 91.69        & 1.30       & 91.46             & 0.19             \\
BYOL           & 91.98         & 0.82        & 91.38         & 0.51        & 91.37        & 1.28       & 90.09             & 0.17             \\ \bottomrule
\end{tabular}
\label{tab:SS_methods}
\end{table}

\section{DBD with Different Self-supervised Methods}
In this paper, we believe that the desired feature extractor is mapping visually similar inputs to similar positions in the feature space, such that poisoned samples will be separated into their source classes. This goal is compatible with that of self-supervised learning. We believe that any self-supervised learning can be adopted in our method. To further verify this point, we replace the adopted SimCLR with other self-supervised methods in our DBD and examine their performance.

\noindent \textbf{Settings.}
We replace the SimCLR with two other self-supervised methods, including MoCo-V2 \citep{chen2020improved} and BYOL \citep{grill2020bootstrap}, in our DBD. Except for the adopted self-supervised method, other settings are the same as those used in Section \ref{sec:main}.

\noindent \textbf{Results.}
As shown in Table \ref{tab:SS_methods}, all DBD variants have similar performances. In other words, our DBD is not sensitive to the selection of self-supervised methods.

\begin{table}[ht]
\centering
\caption{The BA (\%) and ASR (\%) of our DBD defending against four attacks with different label-noise learning methods on CIFAR-10 dataset.}
\begin{tabular}{c|cccccccc}
\toprule
Attack $\rightarrow$         & \multicolumn{2}{c}{BadNets} & \multicolumn{2}{c}{Blended} & \multicolumn{2}{c}{WaNet} & \multicolumn{2}{c}{Label-Consistent} \\ \hline
Method $\downarrow$, Metric $\rightarrow$ & BA            & ASR         & BA            & ASR         & BA           & ASR        & BA                & ASR              \\ \hline
SCE            & 92.41         & 0.96        & 92.18         & 1.73        & 91.20        & 0.39       & 91.45             & 0.34             \\ \hline
GCE            & 92.93         & 0.88        & 93.06         & 1.27        & 92.25        & 1.51       & 91.05             & 0.15             \\
APL            & 92.95         & 1.00        & 92.65         & 0.78        & 92.24        & 1.40       & 91.08             & 0.14             \\ \bottomrule
\end{tabular}
\label{tab:LNL_methods}
\end{table}

\section{DBD with Different Label-noise Learning Methods}

In the main manuscript, we adopt SCE as the label-noise learning method in our second stage. In this section, we explore whether our DBD is still effective if other label-noise methods are adopted.

\noindent \textbf{Settings.}
We replace SCE in our DBD with two other label-noise learning methods, including generalized cross entropy (GCE) \citep{zhang2018generalized} and active passive loss (APL) \citep{ma2020normalized}. Specifically, we adopt the combination of NCE+RCE in APL and use the default hyper-parameters suggested in their original paper. Except for the adopted label-noise learning method, other settings are the same as those used in Section \ref{sec:main}.

\noindent \textbf{Results.}
As shown in Table \ref{tab:LNL_methods}, all DBD variants are effective in reducing backdoor threats ($i.e.$, low ASR) while maintaining high benign accuracy. In other words, our DBD is not sensitive to the selection of label-noise learning methods.

\section{Analyzing Why our DBD is Effective in Defending against Label-Consistent Attack}

In general, the good defense performance of our DBD method against the label-consistent attack (which is one of the clean-label attacks) can be explained from the following aspects:

Firstly, as shown in Figure \ref{fig:motivation}, there is a common observation across different attacks (including both poisoned- and clean-label attacks) that poisoned samples tend to gather together in the feature space learned by the standard supervised learning. The most intuitive idea of our DBD is to prevent such a gathering in the learned feature space, which is implemented by self-supervised learning. As shown in Figure \ref{fig:m_d}, the poisoned samples of label-consistent attack are also separated into different areas in the feature space learned by self-supervised learning. This example gives an intuitive explanation about why our DBD can successfully defend against the label-consistent attack.

Furthermore, it is interesting to explore why the poisoned samples in the label-consistent attack are separated under self-supervised learning since all poisoned samples are from the same target class, rather than from different source classes in poisoned-label attacks. For each poisoned sample in this attack, there are two types of features: the trigger and the benign feature with (untargeted) adversarial perturbations. From the perspective of DNNs, benign samples with (untargeted) adversarial perturbations are similar to samples from different source classes, though these samples look similar from the human's perspective. Thus, it is not surprising that poisoned samples in clean-label attacks can also be separated under self-supervised learning, just like those in poisoned-label attacks.



\end{document}